\documentclass[aps,prd,twocolumn,aps,floatfix,superscriptaddress,nofootinbib]{revtex4-1}

\usepackage{bm}
\usepackage{braket}
\usepackage{color}
\usepackage{dcolumn}
\usepackage{epsfig}
\usepackage[T1]{fontenc}
\usepackage{graphicx}
\usepackage[colorlinks]{hyperref}
\usepackage[latin9]{inputenc}
\usepackage{mathtools}
\usepackage{subcaption}
\usepackage{xcolor}
\usepackage{cleveref}
\usepackage{amssymb}
\usepackage{amsthm}
\usepackage{amsmath}
\begin{document}
	
\title{Quantum Gravity Witness via Entanglement of Masses: Casimir Screening}

\newcommand{\affone}{University of Groningen
	PO Box 72, 9700 Groningen, Netherlands}
\newcommand{\afftwo}{Department of Physics and Astronomy, University College London, Gower Street, WC1E 6BT London, United Kingdom.}
\newcommand{\affthree}{Van Swinderen Institute, University of Groningen, 9747 AG Groningen, The Netherlands.}
\newcommand{\afffour}{Department of Physics, University of Warwick, Gibbet Hill Road, Coventry CV4 7AL, United Kingdom.}

\author{Thomas W. van de Kamp}
\affiliation{\affone}

\author{Ryan J. Marshman}
\affiliation{\afftwo}

\author{Sougato Bose}
\affiliation{\afftwo}

\author{Anupam Mazumdar}
\affiliation{\affthree}

\date{\today}

\begin{abstract}
	{A recently proposed experimental protocol for {\bf Q}uantum {\bf G}ravity induced {\bf E}ntanglement of {\bf M}asses ({\bf QGEM}) requires in principle realizable, but still very ambitious, set of parameters in matter-wave interferometry. Motivated by easing the experimental realization, in this paper, we consider the parameter space allowed by a slightly modified experimental design, which mitigates the Casimir potential between two spherical neutral test-masses by separating the two macroscopic interferometers by a thin conducting plate. Although this set-up will reintroduce a Casimir potential between the conducting plate and the masses, there are several advantages of this design. First, the quantum gravity induced entanglement between the two superposed masses will have no Casimir background. Secondly, the matter-wave interferometry itself will be greatly facilitated by allowing both the mass $10^{-16}-10^{-15}$kg and the superposition size $\Delta x \sim 20 \mu$m to be a one-two order of magnitude smaller than those proposed earlier, and thereby also two orders of magnitude smaller magnetic field gradient of $10^4$Tm$^{-1}$ to create that superposition through the Stern-Gerlach effect. In this context, we will further investigate the collisional decoherences and decoherence due to vibrational modes of the conducting plate.}
\end{abstract}

\maketitle

\section{Introduction}

{\bf Q}uantum {\bf G}ravity induced {\bf E}ntanglement of {\bf M}asses ({\bf QGEM}) is a protocol to test whether gravity is classical or quantum in a table-top experiment \cite{Bose:2017nin,QNG}, and \cite{Marletto2017}. Ref.\cite{QNG} discusses what aspects of the quantum nature of gravity we will test in a lab. The {\bf QGEM} idea is based on a simple fact that an initially pure state of two spatially superposed (quantum) neutral masses will not entangle if they are interacting via a classical gravitational potential. This is in concordance with the ``no generation of entanglement via local operations and classical communication (LOCC)'' theorem~\cite{PhysRevA.59.1070}. As the two superposed quantum masses are prohibited from interacting directly non-locally (action at a distance) because of the local character of quantum field theory (local in comparison to the separation of the masses \cite{QNG}), the ``LO'' part of LOCC cannot be circumvented. Thus the growth of entanglement necessitates circumventing the CC part of LOCC and is thus due to  local 
operations and quantum communication (LOQC). The relevant quantum channel for the QC here has to be an off-shell/virtual graviton~\cite{QNG} in order to maintain a continuous growth of entanglement in a pure state of the two masses according to a $1/r$ interaction~\footnote{The Ref.\cite{QNG} also investigated gravitational theories with non-local interactions. However, in such theories the non-locality has a very mild form within the scale of non-locality. Beyond the scale the theory interpolates to a local theory. Moreover, non-locality does not affect the free theory, but it affects only at the level of interaction~\cite{Biswas:2011ar,Buoninfante:2018mre}., Such non-local effects  may appear in the world line approximation of a string theory ~\cite{Abel:2019zou}.}.

The exchange of a virtual graviton by the two quantum superposed masses is entirely non-classical. In the parlance of quantum field theory,  in a Feynman scattering diagram, the mediator or a Feynman propagator does not obey the Einstein's classical equations of motion or the energy condition. Instead, the virtuality of a graviton is bounded by the energy-time uncertainty relationship, see~\cite{QNG}~\footnote{This proposal has generated much interest in the community, see~\cite{carney2018tabletop, anastopoulos2018comment, hall2018two, marletto2018can, superpositionofgeometries2018, christodoulou2018possibility, belenchia2019information, giampaolo2018entanglement} and paradox resolutions~\cite{carlesso2019testing, al2018optomechanical}. The interpretation of  the experiment through virtual graviton exchange has been provided in Ref.~\cite{QNG}.  There are also tests of ruling out certain models of classical gravity+quantum matter~\cite{Page:1981aj,Ashoorioon:2012kh,carlesso2019testing, al2018optomechanical, Miao2019},\cite{Page:1981aj},\cite{carlesso2019testing},\cite{al2018optomechanical}, which, however, will not unambiguously prove whether gravity ought to quantum in nature, say, as opposed to being simply stochastic, but still classical in nature.}. 

The {\bf QGEM} protocol relies on creating macroscopic spatial superposition of large masses~\cite{Bose:2017nin,QNG}.  Spin is embedded inside the macroscopic masses,  generically a dielectric crystal of micron dimensions. There have been many proposals which have been put forward for creating large mass spatial superpositions~\cite{Bose1997,bose1999scheme,scala2013matter,yin2013large,PhysRevA.84.052121,bateman2014near,marinkovic2018optomechanical, ahn2018optically,  kaltenbaek2016macroscopic,  arndt2014testing, miao2019quantum, krisnanda2019observable,  bykov2019direct, qvarfort2018mesoscopic}. However,  the {\bf QGEM} proposal employs the Stern-Gerlach principle, which has only recently been shown to be viable for interferometry \cite{PhysRevLett.117.143003, Folman2013, Folman2018, folman2019, Morley-Bose2018}, including several ideas for noise reduction in the same \cite{Pedernalles2019, MIMAC2018}.  However, the previously conducted experiments used atoms for the interferometric particles, and superposition sizes and times far below what is required for the proposed experiment~\cite{Bose:2017nin}. This has also led to recent work to determine the tolerable decoherence in the experiment~\cite{nguyen2020entanglement}. Here we will provide an updated scheme for witnessing gravitationally mediated entanglement which employs a perfectly conducting plate to screen electromagnetic interactions between the two masses. In doing so, we will show that the two masses can be made smaller, be placed closer together, and will require a smaller spatial superposition  $\Delta x$ (and hence magnetic field gradient), which will greatly aid in the ease of the implementation of the protocol. We will also consider the sources of decoherence in our set-up. 

This paper also has an additional minor role. While the original QGEM proposal \cite{Bose:2017nin} used considerable simplifications in the details of the analysis in order to keep the conceptual schematic clear, here we systematically optimise the parameter domain for various strengths of the gravitational entangling phase and include also the gravitational phase acquired during the splitting and the recombination parts of the interferometry so as to exploit the full duration of the gravitational interaction. It is only then that we find that in order to alleviate the parameter domain the Casimir screening is necessary. Also note that in the context of a QGEM experiment, Casimir screening is far more non-trivial to investigate in comparison to its usage in accurately estimating close range gravity \cite{Geraci2010} as the introduction of the screen affects the entangling phase and its coherence in a complex way, which we will discuss.

We will begin by describing the original experimental proposal  in Section~\ref{sec:Assumptions} before introducing our new proposed modifications to the set-up Section~\ref{sec:Casimir Screening}. In Sections~\ref{sec:col decoherence} and~\ref{sec: deflection} we conduct a decoherence analysis for the experiment. It will be shown that with this set-up we can use test-masses of $\sim 10^{-15}$ kg and laboratory possible magnetic field gradients of $10^4$ Tm$^{-1}$.


\section[Original set-up and limitations]{Original QGEM set-up \label{sec:Assumptions}}

\begin{figure}[h]
	\includegraphics[width=\columnwidth]{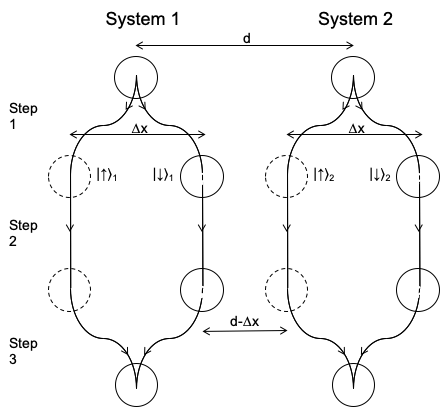}
	\caption{ The {\bf QGEM} set-up comprises two interferometers, system 1 and 2 with the two spatially superposed particles and their trajectories. Their respective spin states have been shown. Note the three paths, Step 1, where the superposition is created by inhomogeneous magnetic field. Step 2, where the test-masses and their superpositions are freely-falling, and Step 3, where the trajectories are brought back for the measurement of spin correlations. The centre of mass distance is $d$ and the superposition is denoted as $\Delta x$. The figure has been adopted from Ref.\cite{Bose:2017nin}, where the details of the pulse sequence that enables the Stern-Gerlach interferometry is also described}. \label{fig:set-up1}
\end{figure}

Here we will discuss the original experimental proposal, {\bf QGEM} \cite{Bose:2017nin}, as shown in Fig.~\ref{fig:set-up1}. In the original paper, we assumed that spins are embedded in the two spherical masses. Let us assume them to be the same material. For the purpose of illustration, here we will consider the system to be diamond with one NV centre point, where the electronic spin can be embedded~\footnote{Note that the diamond may also not be the ideal choice of material. The surface defect of a diamond will modify the classical trajectory in presence of an external inhomogeneous magnetic field, as discussed in \cite{2019arXiv190600835P}. In this paper, we will not consider this effect. We will assume that the surface of the diamond is free from any defects. Furthermore, since our computation relies only on the density of the masses, and not other specific properties, we will assume our test-masses to be perfect spheres with density $\rho=3.5 {\rm gm/cm^3}$.  }.

With the parameters given in Fig.\ref{fig:set-up1}, it can be shown that the entangled wave-function of the combined system of the two test-masses, both with the same mass $m$, {\em after} the completion of the interferometers, when {\em only} that spins embedded in the masses are entangled can be written as \cite{Bose:2017nin}:
\begin{align}
\ket{\Psi(t_{int})}= &\displaystyle \frac{1}{2}e^{i\phi}\bigg[\ket{\uparrow}_1\{\ket{\uparrow}_2+e^{i\Delta\phi_{\uparrow\downarrow}}\ket{\downarrow}_2\} \nonumber\\
+ & \displaystyle \ket{\downarrow}_1\{e^{i\Delta\phi_{\downarrow\uparrow}}\ket{\uparrow}_2+\ket{\downarrow}_2\}\bigg]\,,
\label{eq: wavefunction}
\end{align}
Where $t_{int} $ is the total time of interaction or the total {\it flight time}, the phases are given by: 
\begin{eqnarray}
\phi=\frac{Gm^2}{\hbar d}t_{int},~~\Delta\phi_{\uparrow\downarrow}=\frac{Gm^2}{\hbar (d+\Delta x)}t_{int}-\phi
\nonumber \\
\Delta\phi_{\downarrow\uparrow}=\frac{Gm^2}{\hbar (d-\Delta x)}t_{int} -\phi,
\end{eqnarray}
where $\hbar$ is the reduced Planck constant.  The  corresponding effective entangling phase (i.e., the phase which makes $\ket{\Psi(t_{int})}$ of Eq.(\ref{eq: wavefunction}) an entangled state \cite{Bose:2017nin}, and makes its entanglement as quantified by the von Neumann entropy, which is non-zero~\cite{QNG}), is given by~\footnote{Note that the gravitational potential is the same as that of the classical general relativity. However, for quantum aspects what matters is how the potential arises. If it is a contact potential, then it violates relativity. However, it can be made to satisfy {\em both} relativity and quantum mechanics if the potential arises by an off-shell/virtual  exchange of a graviton; it's origin is then inherently non-classical and quantum~\cite{QNG}.  The quantum induced gravitational potential indeed differentiates from the classical one. The former gives rise to quantum entanglement, while a classical gravity will not lead to any entanglement or increment in the entanglement. The {\bf QGEM} protocol can also test modifications of gravity at short distances~\cite{Biswas:2011ar,Edholm:2016hbt}, see \cite{QNG}.}:
\begin{align}
\Phi_{\text{eff}}= &\displaystyle \Delta\phi_{\uparrow\downarrow} + \Delta\phi_{\downarrow\uparrow} \nonumber\\
= & \displaystyle \frac{Gm^2}{\hbar}t_{int}\left(\frac{1}{d-\Delta x}+\frac{1}{d+\Delta x}-2\frac{1}{d}\right)\,,
\label{eq:entanglement_phase}
\end{align}
We can see that for a fixed $\Phi_{\text{eff}}$, the required mass can be minimised by reducing the distance of closest approach of the two masses, $d-\Delta x$. However, this distance of closest approach has a lower limit for the {\bf QGEM} protocol (Fig.\ref{fig:set-up1}), as the gravitational force is not the only force which can entangle the two systems,  with those of electromagnetic origin competing with it. It is fortunate that there are mitigation methodologies for all but one of the electromagnetic interactions~\footnote{See, for example, Ref.\cite{Bose:2017nin} and its supplemental material lists the techniques for neutralizing the masses, as well as getting rid of the charge multipole-charge multipole interactions, namely by using UV discharge \cite{FMonteiro2020} and physical rotations of the masses \cite{FMonteiro2018} respectively, while the direct magnetic dipole interaction between the two spins is truly negligible in comparison to gravity at the relevant distances.}. However, the entanglement can still form due to the Casimir-Polder force~\cite{Casmir:1947hx} present between the two dielectric spheres -- this was the main hindrance in reducing the distance $d-\Delta x$ to below $200 \mu$m in Ref.\cite{Bose:2017nin}, which, in turn, drove up the needed $\Delta x$ and $m$ (and concomitantly the necessary magnetic field gradient that creates the $\Delta x$). This was also identified in the context of optomechanical experiments for gravitational entanglement~\cite{wan2017quantum} with screening again considered as a potential solution. The Casimir-Polder potential between the two large neutral masses is given by \cite{Casmir:1947hx}, and see also \cite{Bose:2017nin,Kim2005StaticPO,2019arXiv190600835P}
\begin{equation}
V_{CP}\sim-\frac{23\hbar c}{4\pi}\frac{R^6}{r^7}\left(\frac{\epsilon-1}{\epsilon+2}\right)^2\,, \label{eq: Casimir-Polder potential} 
\end{equation}
with $\epsilon$ the dielectric constant of test-masses, $r$ is the separation of two states and $R$ the radius of the corresponding test-masses which are taken to be equal. As the gravitational potential scales as $1/r$, at smaller separations ($d-\Delta x$) the Casimir-Polder force becomes dominant over the gravitational potential. By imposing that the gravitational potential should be at least one order of magnitude larger than the Casimir-Polder potential, such that entanglement due to the electromagnetic-force only has a minimal impact on the effective entanglement phase, implies the bound on the distance ($d-\Delta x$):
\begin{equation}
(d-\Delta x)\geq\left(10\frac{23\hbar c }{4\pi G}\left(\frac{3}{4\pi\rho}\frac{\epsilon-1}{\epsilon+2}\right)^2\right)^{\frac{1}{6}}\approx 157\, \mu m
\label{eq: distance_lower_bound} 
\end{equation}
Where $\rho$ is the density of the test-masses. Defining this separation as a constant, $A\equiv d-\Delta x$, clearly manifests the largest entanglement phase for any given mass is generated at a given distance $A$. This gives us the maximum effective entanglement phase, as
\begin{equation}
\Phi_{\text{eff,max}}=\frac{Gm^2}{\hbar}t_{int}\left(\frac{1}{A}+\frac{1}{2\Delta x \, + \, A}-2\frac{1}{\Delta x\, + \, A}\right)
\label{eq:phase_current_set-up}
\end{equation}
The original paper \cite{Bose:2017nin} proposes the use of a magnetic field gradient $\partial_xB$ in a Stern-Gerlach interferometer to create the spatial superposition \cite{PhysRevLett.117.143003}, giving  
\begin{equation}
\Delta x\sim 2\frac{g\mu_B\partial_xB}{m}\left(\frac{\tau}{2}\right)^2, 
\end{equation}
where $\tau = 500$ ms is the acceleration time and $\partial_x B \sim 10^6$ Tm$^{-1}$ was used. 

However, our approach here is to explore the whole parameter space available to impart a given amount of gravitational phase, and through that, a given amount of given amount of entanglement. Thus we rewrite in terms of the mass needed to generate the entanglement phase and find that to be
\begin{equation}
m=\frac{-3A\Phi_{\text{eff,max}}C-C\sqrt{A\Phi_{\text{eff,max}}(A^3\Phi_{\text{eff,max}}+16DC^2)}}{2(A^3\Phi_{\text{eff,max}}-2DC^2)}
\label{eq:mass_minimum}\,
\end{equation}
where here 
\begin{eqnarray}
C=2g\mu_B\partial_xB\left(\frac{\tau}{2}\right)^2\,,\quad \quad D=\frac{G t_{int}}{\hbar}\,.
\end{eqnarray}
Using the originally proposed interaction time $t_{int}=2.5$ s, one obtains the following required masses: 
an entanglement phase $\Phi_{\text{eff,max}}\sim1$ rad requires a mass of $m\approx 2 \times 10^{-14}$kg, a phase of $\Phi_{\text{eff,max}}\sim0.1$ rad requires a minimum mass of $m \approx 4 \times 10^{-15}$kg and for a phase $\Phi_{\text{eff,max}}\sim0.01$ rad requires a mass of $m \approx 10^{-15}$kg. 

The aim of this paper will be to optimise the experiment with lower magnetic field gradient and smaller mass superposition, such that it will be still feasible to get a detectable gravitationally induced entanglement. Motivated by that let us examine the possibilities of using a superposition of $\Delta x\sim 1\mu$m, which is a full two orders of magnitude smaller than that required in Ref.\cite{Bose:2017nin}, and, in fact, a length for which methodologies other than Stern-Gerlach are also available. Note that in this limit Eq.(\ref{eq:entanglement_phase}) can be recast as:
\begin{equation}
\Phi_{\text{eff,max}}=\frac{2Gt_{int}}{\hbar A^3}\left(\frac{g\mu_B\partial_xB}{2}\,\tau^2\right)^2\,.
\label{eq: constant phase}
\end{equation}
For the originally proposed parameters of the experiment, 
but taking  $\partial_xB\sim 10^4$ Tm$^{-1}$, we can not expect to achieve a higher effective entanglement phase than $\approx 4\times10^{-4}$ rad. Therefore, in order to maximise the entanglement phase along with lowering $\partial_x B$, we would need to alter the original design of the {\bf QGEM}  experiment.


\section[Casimir screening]{Casimir Screening between two superpositions \label{sec:Casimir Screening}}
\subsection{set-up}

\begin{figure}[h]
	\includegraphics[width=\columnwidth]{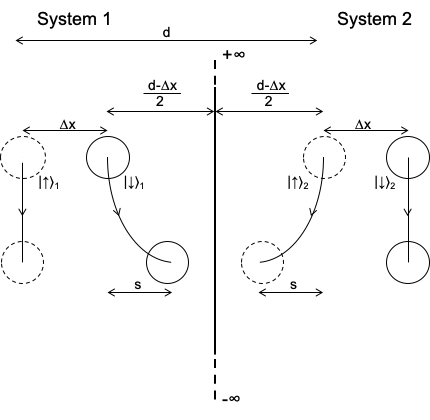}
	\caption{Alternative set-up for step 2 of the \textbf{QGEM} protocol, where we have a perfect conducting plate in between both test-masses. The Casimir force not present between the test-masses allowing for smaller initial separations between the test-masses, generating a higher entanglement phase for a given mass. However, we have to deal with the dynamic distance $s$ caused by the acceleration of the states towards the conducting plate. This displacement $s$ will be controlled principally by the free, dimensionless variable $N$ which sets the initial separation with $d-\Delta x = NR$. \label{fig:set-up2}}
\end{figure}

In order to achieve a detectable entanglement phase we must either employ larger masses and magnetic field gradients, or as we will show here, overcome the limitations on the minimum interaction distance $A$ imposed by the Casimir interaction between the two test-masses. To do this we will consider optimising the minimum distance with a free parameter $N$ given by
\begin{equation}
	A=d-\Delta x=N R \label{eq: Initial separation}
\end{equation}
where $R$ is the radius of the particle. To achieve this, we propose a simple modification to the original set-up by inserting a rigid conducting plate in the middle of two test-masses as depicted in Fig.\ref{fig:set-up2}. We will take this plate to have a metallic, excellent conducting properties  with a thickness $W \sim 1 \mu$m. The {\it key} assumption  here is that the plate is a perfect conductor and assumed to be perfectly reflective. This will screen the electromagnetic interaction between the two superposed masses and act as a {\it Faraday cage}. The Casimir-Polder  potential between the two masses will not be present any more, but there will be Casimir potentials between the conducting plate and the individual superposed masses. Note that this potential between a conducting plate and a sphere is attractive in nature~\cite{Ford:1998ex}.

As a consequence of the Casimir force between the plate and individual masses, we can now allow much smaller separations between the states $\ket{\downarrow}_1$ and $\ket{\uparrow}_2$ than the originally proposed separation $A$, found in Section \ref{sec:Assumptions}. As we will see, this will help us to generate a larger entanglement phase than that given in Eq.(\ref{eq:phase_current_set-up}).  

Since we have to deal with the new Casimir force between the conducting plate and the test states of the superposition itself. This will lead to acceleration of the inner states $\ket{\downarrow}_1$ \& $\ket{\uparrow}_2$ towards the plate during the free-fall of the experiment and introduce a further source of decoherence as we will analyse below. In Fig. \ref{fig:set-up2}, the ``extra'' distance travelled by the inner states is denoted by $s$.

For the purpose of illustration, we consider a square plate to have a length $L=1$ mm and width $W =1 \mu$m. We consider a rigid conducting plate with high density, such as a copper plate $\rho_p=8.96$ g$/$cm$^3$. We will clamp both ends of the plate, and we will let the system free-fall along with the superposed masses of system 1 and 2. We now consider the gravitational attraction between the plate and the masses, and find that it is negligible. The gravitational attraction of the states due to the plate will be given by $a_ g \sim 2\pi G\rho_p W$. This is well justified in the limit when the radius of spherical masses is much smaller than the separation distance between the states and the conducting plate, and the conducting plate can be treated as effectively as infinite in size compared to the test-masses.  The radius  of the sphere is $R\sim 100$ nm. Therefore, the gravitational acceleration of the states with respect to the conducting plate  yields an acceleration $a_g\sim10^{-12}$ m$/$s$^2$. Since this acceleration is negligible, any distance change due to the gravitational interaction due to the presence of the plate can be neglected.


\subsection{Casimir screening}
\label{sec:CS results}

From Ref.~\cite{Ford:1998ex} we find the Casimir force between a sphere and a conducting plate is given by \footnote{Note that the Casimir interaction here is between a mass and a conductive plate, while Eq.(\ref{eq: Casimir-Polder potential}) was between two dielectric spheres. In full, we know that the Casimir interaction between a dielectric sphere and a perfect conducting wall is given by~\cite{Ford:1998ex}:
	\begin{align}
	F_{\text{ca}}= &\displaystyle \frac{1}{4\pi x^4}\int_0^\infty \text{d}\omega\, \alpha_1(\omega)[3\text{sin}(2\omega x)-6\omega x\text{cos}(2\omega x) \nonumber\\
	- & \displaystyle 6\omega^2x^2\text{sin}(2\omega x)+4x^3\omega^3\text{cos}(2\omega x)]\,,
	\label{eq:force_1}
	\end{align}
	where $x$ is the separation distance of the sphere with the closest edge of plate, $\omega$ is  the frequency of the electromagnetic (EM) waves and $\alpha$ is the real part of the static polarizibility, which for a dielectric sphere is $\alpha\sim R^3 \frac{\epsilon(\omega)-1}{\epsilon(\omega)+2}$ \cite{PhysRevLett.117.143003}.
	We have made a reasonable assumption that the test-mass has a constant dielectric function with respect to frequency \cite{article,Floch2011ElectromagneticPO}, and a negligible imaginary component \cite{Ligatchev_2008}.}:
\begin{equation}
F_{\text{ca}}=-\frac{3\hbar c}{2\pi}\left(\frac{\epsilon-1}{\epsilon+2}\right)\frac{R^3}{x^5}.
\label{eq:conductor_casimir_force}
\end{equation}
The above Eq.(\ref{eq:conductor_casimir_force}) shows that the Casimir force becomes infinite for small separations, this implies that there is a certain separation distance at which the acceleration caused by the Casimir force becomes dominant over that caused by the magnetic field, making it impossible to close the interferometers (step 3 of the experiment).  To clarify, if the Casimir force overwhelmed the force due to the magnetic field gradient, it would be impossible to use the magnetic field gradient to bring the two interferometric paths back together. This imposes a new minimum separation distance $d$, such that at the end of the free-fall (step 2), the separation between the inner states and the edge of the plate ($(d-\Delta x)/2-W/2$) is large enough (where $W$ is the width of the conducting plate), so the acceleration caused by the Casimir force is at least one order of magnitude smaller than that due to the magnetic field gradient to bring back the classical path, closing the interferometer. By demanding that:
\begin{equation}
\frac{a_{ca}}{a_{mag}}\leq 0.1 \Rightarrow
x\ge 
\left(\frac{90\hbar c}{8\rho\pi^2}\left(\frac{\epsilon-1}{\epsilon+2}\right)\frac{m}{g\mu_B\partial_xB}\right)^{1/5}\,,
\label{eq:acc requirement}
\end{equation}
during the free-fall, the inner states will travel a distance $s$ inwards during the  time interval $\Delta t$.
Note that the centre of mass distance $d$ is a ``tunable'' distance in an experiment, controlled by the parameter $N$ as given in Eq.~\ref{eq: Initial separation}. During this free-fall the effective entanglement phase of Eq.(\ref{eq:entanglement_phase}) becomes for an infinitesimal time period $\Delta t$
\begin{align}
\Phi_{\text{eff}}=&\frac{Gm^2}{\hbar}\Delta t \bigg(\frac{1}{NR-2s} \nonumber\\
&+\frac{1}{2\Delta x +NR}-2\frac{1}{\Delta x +NR-s}\bigg).
\label{eq: entanglement phase alternative}
\end{align}
%

\begin{figure}[h]
	\includegraphics[width=\columnwidth]{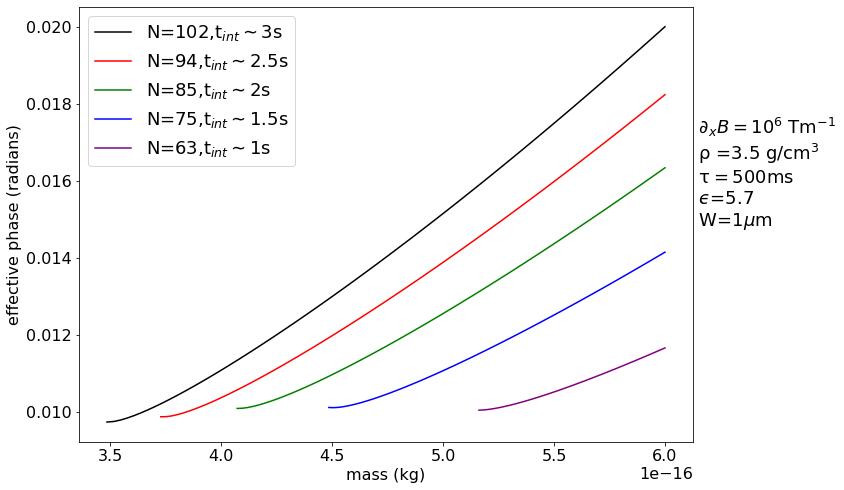}
	\caption{The plot shows the maximum entanglement phase generated during step 2 using the alternative set-up of the {\bf QGEM} protocol as a function of mass for different initial separations, characterised by the parameter $N$. The lowest mass indicates the minimum mass for which the acceleration requirement  Eq.(\ref{eq:acc requirement}) is barely met. The values of $\partial_x B, \rho, \tau_{acc},\epsilon, W$ are labelled above. \label{fig:dynamic01}}
\end{figure}

%
By computing this phase for different initial separations, dictated here by the parameter $N$, we can optimise the phase such that the bound given in Eq.(\ref{eq:acc requirement}) is saturated at the end of the free-fall for a given mass, $m$ ,while maximising the phase $\Phi_{\text{eff}}$.  Here we show the results in Fig.\ref{fig:dynamic01}, which displays the optimum results, provided the entanglement phase of order $\Phi_{\rm eff}=0.01$  rad, for different interaction times denoted here by $N$.

Note that the acceleration due to the Casimir potential is independent of the size of the test-masses, see Eq. (\ref{eq:conductor_casimir_force}). However $d$ can be chosen to scale with the radius $R$ for the purpose of optimisation of the phase and the Casimir potential between systems 1 and 2.
Following this, our  Eq.(\ref{eq:acc requirement}) would then scale as $R^{3/5}$, whereas our dynamic distance scales with $R$. Therefore, by fixing the value $N$, we will find the minimum mass for which the condition  Eq.(\ref{eq:acc requirement}) is barely met.  Masses below this will lead to a violation of the bound set by Eq.(\ref{eq:acc requirement}). 
The resulting minimum mass correspond to the lowest mass seen in Fig.\ref{fig:dynamic01}. These masses coincide with the chosen minimum final entanglement phase requirement, which we have set  to be $\Phi_{\text{eff}}=0.01$ rad. If we consider 
$t_{int}\sim 2.5$s, roughly the total flight time, we note that a mass of approximately $m \sim 3.7\times10^{-16}$ kg is sufficient to saturate the entanglement phase~\footnote{ In fact, an interaction time of only $t_{int}\sim 1$s already allows for a $50\%$ reduction to the required mass compared to the original set-up where $m\sim 10^{-14}$Kg. We can also conduct a similar analysis for a phase requirement $\Phi_{\text{eff}}=1$ rad to find that, for an interaction time of $t_{int}=2.5$ s, a mass of $3.8\times10^{-15}$ kg is adequate.}.

\begin{figure}[h]
	\includegraphics[width=\columnwidth]{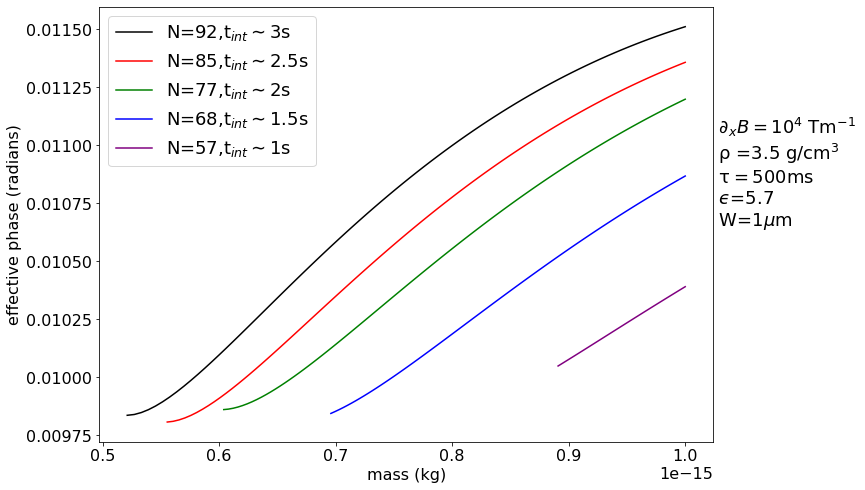}
	\caption{The plot shows the maximum entanglement phase generated during step 2 using the alternative set-up of the {\bf QGEM} protocol as a function of mass for different initial separations, characterised by the parameter $N$. The lowest mass indicates the minimum mass for which the acceleration requirement  given in Eq.(\ref{eq:acc requirement}) is barely met. The values of $\partial_x B, \rho, \tau_{acc},\epsilon, W$ are labelled above.  \label{fig:dynamic0100}}
\end{figure}

Similar analysis can be performed by reducing the required magnetic field gradient to $\partial_xB=10^4$ Tm$^{-1}$, which is much more feasible with current technology. Following the same procedure as before, and requiring  $\Phi_{\text{eff}}=0.01$,  we obtain the allowed parameter space as given in Fig. \ref{fig:dynamic0100}.

The figs. \ref{fig:dynamic01} and \ref{fig:dynamic0100}  show that by inserting a conducting plate, the magnetic field gradient can be reduced by two orders of magnitude, which also reduces the mass of macroscopic superposition by one or two orders of magnitude compared to the original set-up~\cite{Bose:2017nin}~\footnote{Note that in the above analysis we have ignored the outer states from the conducting plate of figure \ref{fig:set-up2}.  As we will point out here that since the Casimir potential drops as $x^{-5}$, the change in $\Delta x$ is truly negligible. The positional displace is of the order of $10^{-3}R\sim 10^{-1}$nm after an interaction time of $2.5$ s, which is truly negligible compared to $\Delta x\sim 10\mu$m.}. Specifically we can use a magnetic field gradient of $10^{4}$ Tm$^{-1}$ and a flight time of only $1$ s by placing the masses closer together with an initial separation of $47$ $\mu$m (corresponding to $N=57$), a initial superposition size $\Delta x = 23$ $\mu$m which grows by a further $s=2$ $\mu$m using lighter masses ($m\sim 10^{-15}$ kg). 

\section{Entanglement Witness \label{sec:witness}}

For the rest of the analysis we will shift our focus towards the most reasonable set of system parameters, and use an interaction time of $t_{int}=1$s. This reduced interaction time reduces the run time of the experiment significantly. As we want to allow the magnetic field gradient of $10^4$ Tm$^{-1}$, we can use a mass of around $10^{-15}$ kg  (see Fig.\ref{fig:dynamic0100}). This will result in an effective entanglement phase $\Phi_{\text{eff}}\sim 0.01$ rad for $N=57$ accrued during step 2. However, there is also a significant phase evolution during step 1 and step 3 of the experiment. The accumulated phase during step 1 and step 3 are described in appendix \ref{sec:extra phase}, and will result in the the total accumulated phase becomes:
\begin{equation}
\Phi_{\text{eff}}\sim 0.015
\label{eq: Phi value}
\end{equation}
Note that an entanglement witness ${\cal W}=I^{(1)} I^{(2)}-\sigma_x^{(1)}\sigma_x^{(2)}-\sigma_y^{(1)}\sigma_z^{(2)}-\sigma_x^{(1)}\sigma_z^{(2)}$ is able detect entanglement of the test-masses provided $\textrm{Tr}(\mathcal{W}\rho)<0$, with $\rho$ the reduced density matrix of the entangled test-masses~\footnote{This form of $\mathcal{W}$ was found, in Ref.\cite{Chevalier:2020uvv}, to be a more suitable witness for the original experiment \cite{Bose:2017nin}.}. 
The decoherence can be modelled as a reduction of the off-diagonal elements of the density matrix. This decoherence is present throughout the whole runtime of steps 1 to 3 of the experiment, this means that in order for $Tr(\mathcal{W}\rho)<0$ we have the requirement \cite{Chevalier:2020uvv}

\begin{equation}
\gamma t_{int} < \frac{\Phi_{\textrm{eff}}}{2}
\end{equation}
With $\gamma$ the decoherence rate. This implies that for our accumulated phase of Eq.(\ref{eq: Phi value}), a bound on  $\gamma t$< $\Phi_{\text{eff}}/2\sim 0.0075$ is required. In the following sections we will discuss various sources of decoherence.

\section{Collisional decoherence}
\label{sec:col decoherence}

One of the primary sources of decoherence for a macroscopic spatial superposition will be due to the air molecules present in the vacuum chamber, and the absorption and emission of blackbody photons, see 
\cite{decoherence, PhysRevA.84.052121}.In this section we will analyse these decoherence rates with a test-mass of $m=10^{-15}$kg, $N=57$ and an interaction time of $1$ s for the purpose of illustration, but we can do a similar analysis for other possibilities.

\begin{figure}[h]
	\includegraphics[width=0.89\columnwidth]{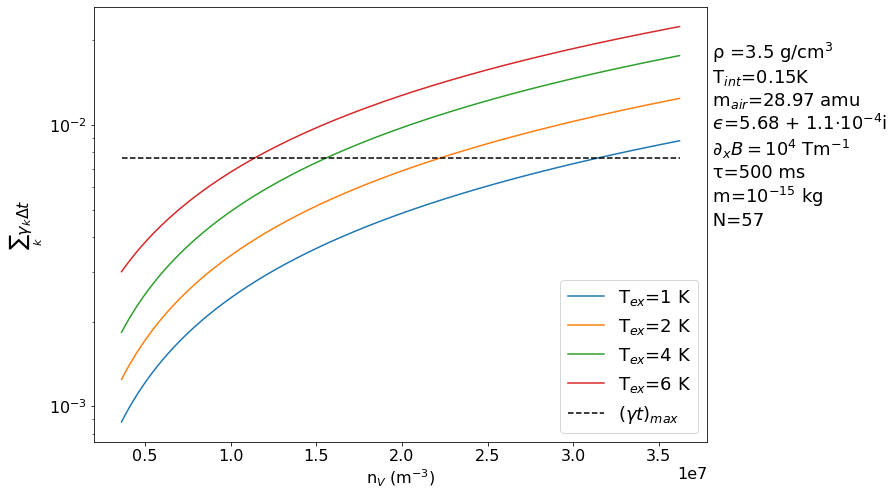}
	\caption{Decoherence of a single superposition as a function of the number density of the environmental gas for different external temperatures computed with  Eq.(\ref{eq:full decohere analytic2}) together with the maximum allowed decoherence factor (black dotted line) for measuring entanglement for the parameters giving next to the graph. Note  that the actual decoherence rate is exp$[-\sum\gamma_k\Delta t]$.\label{fig:decohererate2}}
\end{figure}

The largest superposition size, $\Delta x +s_{max}\sim10^{-5}$ m, will occur at the end of the free-fall, end of step-2. This is smaller than the thermal wavelength of blackbody photons $\lambda_{bb}\sim 10^{-3}$m for the external ambience temperature of $T_{ex}\sim {\cal O}(1)$k~\footnote{Note that here we have to consider the internal temperature, $T_{int}$, of the test-mass as well. For diamond below $T_{int} < 4$K, the phonon excitations are negligible, see \cite{Bose:2017nin}. }. This implies that the long wavelength approximation for decoherence, see  \cite{decoherence}, is applicable as the superposition size is much smaller than the blackbody photons. 
However, the thermal wavelength of air molecules in the vacuum at these temperatures are in the order of $\lambda_{air}\sim 10^{-10}$ m.Therefore, for the air molecule it is fairly reasonable to assume the short-wavelength limit for computing the decoherence rate, see \cite{decoherence}, during the whole duration of the experiment. 
By applying these limits, we will obtain  the total decoherence factor to be
\begin{equation}
\text{exp}[-\gamma t]=\text{exp}[-(\Gamma_{air}t+\sum_i^3\Lambda_i\sum_k^{3}(x_{\ket{\uparrow}_k}-x_{\ket{\downarrow}_k})^2)\Delta t ] \label{eq:decoherence factor}
\end{equation}
Here $\Gamma_{\text{air}}$ is the scattering rate of the ambient air molecules inside the vacuum chamber, $\Lambda_{i}$'s are the scattering constant of the scattered blackbody photons, the absorbed photons, and the emitted photons from the test-mass. The form ($x_{\ket{\uparrow}_k}-x_{\ket{\downarrow}_k}$) denotes the distance between the superposition states inside a single interferometer at a time $k\Delta t$, with k an integer, which varies throughout the experiment. See appendix \ref{sec: scattering constants} for more details.  The final expression for the total decoherence factor during the whole duration of the experiment becomes (as shown in appendix \ref{sec:decoherence rate})~\footnote{Note that the coherence of the spin state inside diamond will also suffer during the spin flips, however, this effect we are not taking into account here.} :
\begin{widetext}
	\begin{equation}
	\sum_k\gamma_k\Delta t=  
	\left [\Gamma_{\text{air}}(t_{\text{int}}+\tau+\tau_1 ) 
	+ 
	\sum_{i=1}^3\Lambda_i\bigg(\frac{46}{15}a_{mag}^2\{\left(\frac{\tau}{2}\right)^5 +\left(\frac{\tau_1}{2}\right)^5\}+4a_{mag}^2\left(\frac{\tau}{2}\right)^4t_{int} 
	+ 
	\sum_k(4a_{mag}\left(\frac{\tau}{2}\right)^2 s_k+s_k^2)\Delta t\bigg)\right ]
	\label{eq:full decohere analytic2}
	\end{equation}
\end{widetext}
Here $\tau_1=2\sqrt{\left(\frac{\tau}{2}\right)^2 +\frac{s_{\text{max}}}{a_{mag}}}$ and 
$a_{mag}$ is the acceleration of the magnetic field, and $t_{int}$ is the total interaction time, including all the three steps.
In Fig.\ref{fig:decohererate2}, we show how $\sum_k\gamma_k\Delta t$ evolves with the number density $n_V$ of the air molecules inside the vacuum chamber.

We found that the number density of the air molecule should be low, $n_V\sim10^7$ m$^{-3}$. For these low number densities ideal gas law holds. These number densities then correspond to the vacuum pressure $P\sim 5\times 10^{-16}$  which is slightly lower than the proposed pressure of $10^{-15}$ Pa in Ref.\cite{Bose:2017nin}. 

Further note that for the parameters we have shown in the plot, the decoherence due to scattering of air molecules is the dominant source of decoherence as shown by Eqs.\ref{eq: scatter constant air} and \ref{eq:Lambda_s_e_a}. Since, this is independent of the internal temperature of the test-mass, we can further increase this temperature to a few Kelvin  without interfering with the results significantly. 


\section{deflection of the plate}
\label{sec: deflection}

Interactions between the two masses and the conductive plate will excite the vibrational modes of the plate. A small uncertainty in the initial  displacement of the test-masses will lead to an unknown net force acting on both sides of the plate in Fig.\ref{fig:set-up2}. Here we will demonstrate that under certain conditions, these vibrational modes can be minimised. Any differences in the vibrational states of the plate for different positions of the pair of masses is going to decohere the combined state of the masses.

This deflection of the conducting plate is maximum if the force acts on the centre of the plate as it is clamped at either end. As such we will consider this worse case scenario, a point force acting on the middle of the plate. The point source approximation here is good as the the length of the plate exceeds the radius of the two masses (assumed to be perfectly spherical). Note that we have set the coordinates such that along the length of the plate it is $z$-axis and the $y$-axis is onto-the plate while $x$-axis denotes the distance of the test-masses from the plate.  For a plate clamped along the z-axis, we get a deflection $\delta_d$ of the center of the plate due to imbalance Casimir force of \cite{book:2261446}
\begin{equation}
\delta_d=\frac{F_{ca}L^3}{192EI_y}=\frac{FL^2}{16EW^3}
\label{eq:displacement}
\end{equation}

Where $E$ is the Young's modulus of the plate, $I_y$ is the moment of area in the xy-plane of the plate. We assume a square plate of length $L$ and the thickness $W$. For the external temperature $T\sim {\cal O}(1)$ K, and taking the conductor to be copper, the value of Young's modulus of the plate is $E=137$GPa. 

\begin{figure}
	\includegraphics[width=0.89\columnwidth]{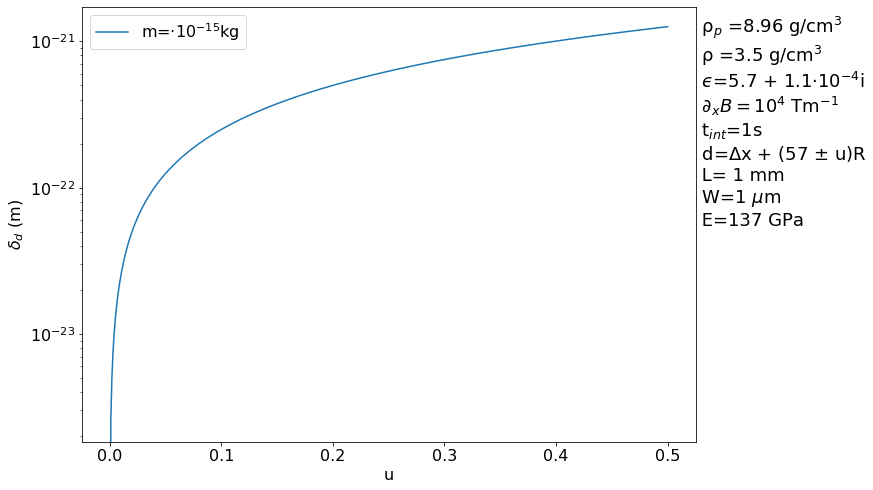}
	\caption{Deflection of the conducting plate, which is assumed to be made out of copper, for an range of displacement errors of the test-masses to the right of figure \ref{fig:set-up2} in terms of the radius of the test-mass for the parameters given next to the graph. \label{fig:deflection}}
\end{figure}

Let us now denote the uncertainty in the placement of a single test-mass relative to the conducting plate by $uR$, where $R$ is the radius of the test-mass, and $0< u< 0.5$. In our set-up there are pair wise test-masses for systems 1 and 2. Therefore, the maximum force imbalance arises when both the test-masses are displaced by a distance $uR$  to one side of the $x$-axis. Since the main force is due to Casimir here, we can maximise this force up to the point of free-fall (end of step-2) for both the systems 1 and 2, and compute the maximum deflection to be:
\begin{equation}
\delta_{d,max}=\frac{F_{max}L^2}{16EW^3}=\frac{|\textbf{F}_{\ket{\downarrow}_1}(x(t_{int}))+\textbf{F}_{\ket{\uparrow}_2}(x(t_{int}))|L^2}{16EW^3}
\label{eq:delta_d max}
\end{equation}
In Fig.\ref{fig:deflection} we show $\delta_d$ with respect to the uncertainty in the initial displacement of the test-masses from the plate $0< u <0.5$. observe that the deflection is truly negligible compared to the thickness of the plate, implying that one doesn't have to worry about the deflection of the plate. Furthermore, if we consider the plates displacement due to the $|\uparrow\rangle_{1}|\uparrow\rangle_{2}$ and $|\downarrow\rangle_{1}|\downarrow\rangle_{2}$ states, we found that, there is still a negligible deflection bounded from above by $\sim 5\times 10^{-21}$m. This is as we sill see shortly orders of magnitudes smaller than the ground state of the plate, such that no which path information is imprinted onto the plate, as such no significant decoherence is expected.

From Eq.(\ref{eq:delta_d max}) we can deduce the corresponding frequency of oscillation of the whole plate to be:
\begin{equation}
\omega=\sqrt{\frac{k}{m}}=\sqrt{\frac{16EW^3}{mL^2}}=\sqrt{\frac{16EW^2}{\rho_p L^4}}\,,
\end{equation}
where $m$ is the mass of the plate and $\rho_p$ is the density of the plate. We can now compute the ground state spread of the plate to be roughly given by:
\begin{equation}
\Delta S=\sqrt{\frac{\hbar}{m\omega}}=\frac{1}{W}
\sqrt{\frac{\hbar}{\rho\sqrt{16E}}}\,.
\end{equation}
Therefore, by demanding that the plate itself  does not encode significant which path information due to the different Casimir force imbalances for each of the possible positions of the masses, this ground state quantum spread should be larger than the deflection of the plate, $\delta_d$. This will further constrain the mass of the plate. In our case we can select the length of the plate to be a free parameter, which is then constrained as
\begin{equation}
L < \left(\frac{1}{W^2}\frac{|\textbf{F}_{\ket{\downarrow}_1}(x(t_{int}))+\textbf{F}_{\ket{\uparrow}_2}(x(t_{int}))|}{(16E)^{3/4}}\sqrt{\frac{\rho}{\hbar}}\right)^{-1/2}
\end{equation}
For the optimal parameters of the experiment, $t_{int}=1$s,~$N=57$, $\partial_x B=10^4$ Tm$^{-1}$, $m=10^{-15}$kg, 
and $u=\pm 0.5R$, we find that  $L < 100$mm. This implies that for $L=1$mm is suitable, the oscillations in the plate induced by the Casimir force will not spoil the coherence of the test-masses.

\section{Other considerations}
While we have alleviated one specific problem and not necessarily solved all the difficulties of {\bf QGEM} we still point out a few potential ways to address them. Knowing/controlling the fluctuations in the position of each mass is a hurdle of the original {\bf QGEM} protocol, which is not specifically addressed by the Casimir screening idea, whose purpose is electromagnetic shielding.   However, in exciting recent developments, levitated masses can now be cooled to their motional ground states \cite{Delic892, PhysRevLett.123.153601} so that  thermal fluctuations of motion are not an issue any more. However, motional fluctuations can also stem from random nuclear spins in the mass, as even for a $>99.9995 \%$ isotopically pure sample~\cite{dwyer2014enriching} (be it a silicon crystal or a diamond crystal or some other dielectric) on average a total nuclear moment equivalent to at most 1 random electronic spin moment could be present in each $\sim 10^{-15}$Kg mass. This random moment is quasi-static -- fluctuates only from run to run of the experiment. Thus one can gently modify the scheme of Ref.\cite{Bose:2017nin} so as to incorporate frequent reversals of {\em both} the magnetic field gradient and the electronic spin in tandem in the splitting and recombination stages, while the nuclear spin moment remains unaffected. Such frequent magnetic field direction reversals also have the potential to cancel the contribution of the diamagnetic energy to the Hamiltonian of the system (this has been pointed out to be an important term in the Hamiltonian in Ref.\cite{Pedernalles2019}) when integrated over the full time scale of the splitting/recombination, while the electron-motion coupling terms remains unchanged as the electronic spin is flipped each time the magnetic field changes direction.  As pointed out in \cite{MIMAC2018} the magnetic field sources of such interferometers would be ``shaped'' magnets or current carrying wires., which essentially shift the centre of the diamagnetic trap according to the path so that the superposition size is not limited by the diamagnetism either. Additionally, if the masses used are composite materials such as nickel coated diamond (thin layer of nickel on diamond), the diamagnetic core's effect can be cancelled by the ferromagnetic coating of very small appropriate thickness. Thus we believe that the other challenges in {\bf QGEM} are not insurmountable, and several techniques are potentially available, although their detailed studies are beyond the scope of this paper.

\section{Conclusion}
The original {\bf QGEM} proposal required a mechanism to overcome the Casimir-Polder potential. Here we have provided a simple solution; inserting a conducting plate between the two quantum superpositions, denoted here by system 1 and 2, see Fig.\ref{fig:set-up2}. The conducting plate screens electromagnetic interactions and Casimir potential between the two superposed neutral masses allowing for smaller separations between the neutral masses. Doing so, provides many exciting outcomes. Namely, the particle masses can be reduced to around $10^{-15}-10^{-16}$kg. However, the conducting plate introduces a Casimir force between the two masses and the plate itself. The force is attractive and tends to modify the trajectories of the particles in such a way that the initial  size of spatial superposition can now be made slightly smaller. A smaller mass and a smaller spatial superposition can be achieved with the current state of strength for the inhomogeneous magnetic field, i.e. of order $10^{4}{\rm Tm^{-1}}$. We have also found that the flight time of the experiment can also be reduced to roughly $1$ s. A smaller mass and the reduced flight time will yield a lower, but still detectable  entanglement phase of order $0.01$ rad. 

We have also analysed various sources of decoherence in the experiment. We have considered collisional decoherence due to air molecules and blackbody radiation. We have also analysed a new source of decoherence due to Casimir potential between the test-mass and the conducting plate. We have found that it is possible to mitigate the decoherence due to vibrational motion of the conducting plate, provided the plate is relatively rigid, such as a copper square plate with a $1$ mm length and height and a thickness $1\mu$m.

Although, our results highlight a small improvement in some of the experimental parameters, step by step it brings us closer towards  entangling two macroscopic quantum superpositions in a table-top experiment. In fact, it is now eminently possible with  magnetic field gradients which is well within the reach of current generation laboratories. Tests such as {\bf QGEM} are essential for constructing sensible quantum theory of gravity at all energies. The gravity remains the only interactions of nature whose quantum properties are not known in any experiment. Although, the concept of graviton is viewed as a consequence of a perturbative treatment of quantum gravity, nevertheless, it is a vital tool for non-perturbative aspects of quantum gravity as well, such as strings~\cite{Biswas:2011ar,Abel:2019zou}. Moreover, the {\bf QGEM} protocol highlights that if gravity is quantum, it would inevitably entangle matter. This will have implications of quantum gravity at all energies and can be considered as a complementary path of research from AdS/CFT correspondence~\cite{Maldacena:1997re}.

\section{Acknowledgements}
We would like to thank   Prachi Parashar, K. V. Shajesh and particularly M. B. Plenio for extremely insightful discussions on Casimir effects and other potential noise backgrounds in the proposed setups. AM  would like to thank George Palasantzas and Vitaly Svetovoy for very exciting discussions on Casimir-Polder  potential. A.M. is supported by Netherlands Organisation for Scientific Research (NWO) grant no. 680-91-119.  S. B. would like to acknowledge EPSRC grants No. EP/N031105/1 and EP/S000267/1. R. J. M. is supported by a UCL departmental studentship. 

\bibliography{bibliography}

\begin{thebibliography}{61}%
\makeatletter
\providecommand \@ifxundefined [1]{%
 \@ifx{#1\undefined}
}%
\providecommand \@ifnum [1]{%
 \ifnum #1\expandafter \@firstoftwo
 \else \expandafter \@secondoftwo
 \fi
}%
\providecommand \@ifx [1]{%
 \ifx #1\expandafter \@firstoftwo
 \else \expandafter \@secondoftwo
 \fi
}%
\providecommand \natexlab [1]{#1}%
\providecommand \enquote  [1]{``#1''}%
\providecommand \bibnamefont  [1]{#1}%
\providecommand \bibfnamefont [1]{#1}%
\providecommand \citenamefont [1]{#1}%
\providecommand \href@noop [0]{\@secondoftwo}%
\providecommand \href [0]{\begingroup \@sanitize@url \@href}%
\providecommand \@href[1]{\@@startlink{#1}\@@href}%
\providecommand \@@href[1]{\endgroup#1\@@endlink}%
\providecommand \@sanitize@url [0]{\catcode `\\12\catcode `\$12\catcode
  `\&12\catcode `\#12\catcode `\^12\catcode `\_12\catcode `\%12\relax}%
\providecommand \@@startlink[1]{}%
\providecommand \@@endlink[0]{}%
\providecommand \url  [0]{\begingroup\@sanitize@url \@url }%
\providecommand \@url [1]{\endgroup\@href {#1}{\urlprefix }}%
\providecommand \urlprefix  [0]{URL }%
\providecommand \Eprint [0]{\href }%
\providecommand \doibase [0]{http://dx.doi.org/}%
\providecommand \selectlanguage [0]{\@gobble}%
\providecommand \bibinfo  [0]{\@secondoftwo}%
\providecommand \bibfield  [0]{\@secondoftwo}%
\providecommand \translation [1]{[#1]}%
\providecommand \BibitemOpen [0]{}%
\providecommand \bibitemStop [0]{}%
\providecommand \bibitemNoStop [0]{.\EOS\space}%
\providecommand \EOS [0]{\spacefactor3000\relax}%
\providecommand \BibitemShut  [1]{\csname bibitem#1\endcsname}%
\let\auto@bib@innerbib\@empty
\bibitem [{\citenamefont {Bose}\ \emph {et~al.}(2017)\citenamefont {Bose},
  \citenamefont {Mazumdar}, \citenamefont {Morley}, \citenamefont {Ulbricht},
  \citenamefont {Toro\'~s}, \citenamefont {Paternostro}, \citenamefont
  {Geraci}, \citenamefont {Barker}, \citenamefont {Kim},\ and\ \citenamefont
  {Milburn}}]{Bose:2017nin}%
  \BibitemOpen
  \bibfield  {author} {\bibinfo {author} {\bibfnamefont {S.}~\bibnamefont
  {Bose}}, \bibinfo {author} {\bibfnamefont {A.}~\bibnamefont {Mazumdar}},
  \bibinfo {author} {\bibfnamefont {G.~W.}\ \bibnamefont {Morley}}, \bibinfo
  {author} {\bibfnamefont {H.}~\bibnamefont {Ulbricht}}, \bibinfo {author}
  {\bibfnamefont {M.}~\bibnamefont {Toro\'~s}}, \bibinfo {author}
  {\bibfnamefont {M.}~\bibnamefont {Paternostro}}, \bibinfo {author}
  {\bibfnamefont {A.}~\bibnamefont {Geraci}}, \bibinfo {author} {\bibfnamefont
  {P.}~\bibnamefont {Barker}}, \bibinfo {author} {\bibfnamefont
  {M.}~\bibnamefont {Kim}}, \ and\ \bibinfo {author} {\bibfnamefont
  {G.}~\bibnamefont {Milburn}},\ }\href {\doibase
  10.1103/PhysRevLett.119.240401} {\bibfield  {journal} {\bibinfo  {journal}
  {Phys. Rev. Lett.}\ }\textbf {\bibinfo {volume} {119}},\ \bibinfo {pages}
  {240401} (\bibinfo {year} {2017})},\ \Eprint
  {http://arxiv.org/abs/1707.06050} {arXiv:1707.06050 [quant-ph]} \BibitemShut
  {NoStop}%
\bibitem [{\citenamefont {Marshman}\ \emph {et~al.}(2020)\citenamefont
  {Marshman}, \citenamefont {Mazumdar},\ and\ \citenamefont {Bose}}]{QNG}%
  \BibitemOpen
  \bibfield  {author} {\bibinfo {author} {\bibfnamefont {R.~J.}\ \bibnamefont
  {Marshman}}, \bibinfo {author} {\bibfnamefont {A.}~\bibnamefont {Mazumdar}},
  \ and\ \bibinfo {author} {\bibfnamefont {S.}~\bibnamefont {Bose}},\ }\href
  {\doibase 10.1103/PhysRevA.101.052110} {\bibfield  {journal} {\bibinfo
  {journal} {Phys. Rev. A}\ }\textbf {\bibinfo {volume} {101}},\ \bibinfo
  {pages} {052110} (\bibinfo {year} {2020})}\BibitemShut {NoStop}%
\bibitem [{\citenamefont {Marletto}\ and\ \citenamefont
  {Vedral}(2017)}]{Marletto2017}%
  \BibitemOpen
  \bibfield  {author} {\bibinfo {author} {\bibfnamefont {C.}~\bibnamefont
  {Marletto}}\ and\ \bibinfo {author} {\bibfnamefont {V.}~\bibnamefont
  {Vedral}},\ }\href {\doibase 10.1103/PhysRevLett.119.240402} {\bibfield
  {journal} {\bibinfo  {journal} {Phys. Rev. Lett.}\ }\textbf {\bibinfo
  {volume} {119}},\ \bibinfo {pages} {240402} (\bibinfo {year}
  {2017})}\BibitemShut {NoStop}%
\bibitem [{\citenamefont {Bennett}\ \emph {et~al.}(1999)\citenamefont
  {Bennett}, \citenamefont {DiVincenzo}, \citenamefont {Fuchs}, \citenamefont
  {Mor}, \citenamefont {Rains}, \citenamefont {Shor}, \citenamefont {Smolin},\
  and\ \citenamefont {Wootters}}]{PhysRevA.59.1070}%
  \BibitemOpen
  \bibfield  {author} {\bibinfo {author} {\bibfnamefont {C.~H.}\ \bibnamefont
  {Bennett}}, \bibinfo {author} {\bibfnamefont {D.~P.}\ \bibnamefont
  {DiVincenzo}}, \bibinfo {author} {\bibfnamefont {C.~A.}\ \bibnamefont
  {Fuchs}}, \bibinfo {author} {\bibfnamefont {T.}~\bibnamefont {Mor}}, \bibinfo
  {author} {\bibfnamefont {E.}~\bibnamefont {Rains}}, \bibinfo {author}
  {\bibfnamefont {P.~W.}\ \bibnamefont {Shor}}, \bibinfo {author}
  {\bibfnamefont {J.~A.}\ \bibnamefont {Smolin}}, \ and\ \bibinfo {author}
  {\bibfnamefont {W.~K.}\ \bibnamefont {Wootters}},\ }\href {\doibase
  10.1103/PhysRevA.59.1070} {\bibfield  {journal} {\bibinfo  {journal} {Phys.
  Rev. A}\ }\textbf {\bibinfo {volume} {59}},\ \bibinfo {pages} {1070}
  (\bibinfo {year} {1999})}\BibitemShut {NoStop}%
\bibitem [{\citenamefont {Biswas}\ \emph {et~al.}(2012)\citenamefont {Biswas},
  \citenamefont {Gerwick}, \citenamefont {Koivisto},\ and\ \citenamefont
  {Mazumdar}}]{Biswas:2011ar}%
  \BibitemOpen
  \bibfield  {author} {\bibinfo {author} {\bibfnamefont {T.}~\bibnamefont
  {Biswas}}, \bibinfo {author} {\bibfnamefont {E.}~\bibnamefont {Gerwick}},
  \bibinfo {author} {\bibfnamefont {T.}~\bibnamefont {Koivisto}}, \ and\
  \bibinfo {author} {\bibfnamefont {A.}~\bibnamefont {Mazumdar}},\ }\href
  {\doibase 10.1103/PhysRevLett.108.031101} {\bibfield  {journal} {\bibinfo
  {journal} {Phys. Rev. Lett.}\ }\textbf {\bibinfo {volume} {108}},\ \bibinfo
  {pages} {031101} (\bibinfo {year} {2012})},\ \Eprint
  {http://arxiv.org/abs/1110.5249} {arXiv:1110.5249 [gr-qc]} \BibitemShut
  {NoStop}%
\bibitem [{\citenamefont {Buoninfante}\ \emph {et~al.}(2019)\citenamefont
  {Buoninfante}, \citenamefont {Lambiase},\ and\ \citenamefont
  {Mazumdar}}]{Buoninfante:2018mre}%
  \BibitemOpen
  \bibfield  {author} {\bibinfo {author} {\bibfnamefont {L.}~\bibnamefont
  {Buoninfante}}, \bibinfo {author} {\bibfnamefont {G.}~\bibnamefont
  {Lambiase}}, \ and\ \bibinfo {author} {\bibfnamefont {A.}~\bibnamefont
  {Mazumdar}},\ }\href {\doibase 10.1016/j.nuclphysb.2019.114646} {\bibfield
  {journal} {\bibinfo  {journal} {Nucl. Phys. B}\ }\textbf {\bibinfo {volume}
  {944}},\ \bibinfo {pages} {114646} (\bibinfo {year} {2019})},\ \Eprint
  {http://arxiv.org/abs/1805.03559} {arXiv:1805.03559 [hep-th]} \BibitemShut
  {NoStop}%
\bibitem [{\citenamefont {Abel}\ \emph {et~al.}(2020)\citenamefont {Abel},
  \citenamefont {Buoninfante},\ and\ \citenamefont {Mazumdar}}]{Abel:2019zou}%
  \BibitemOpen
  \bibfield  {author} {\bibinfo {author} {\bibfnamefont {S.}~\bibnamefont
  {Abel}}, \bibinfo {author} {\bibfnamefont {L.}~\bibnamefont {Buoninfante}}, \
  and\ \bibinfo {author} {\bibfnamefont {A.}~\bibnamefont {Mazumdar}},\ }\href
  {\doibase 10.1007/JHEP01(2020)003} {\bibfield  {journal} {\bibinfo  {journal}
  {JHEP}\ }\textbf {\bibinfo {volume} {01}},\ \bibinfo {pages} {003} (\bibinfo
  {year} {2020})},\ \Eprint {http://arxiv.org/abs/1911.06697} {arXiv:1911.06697
  [hep-th]} \BibitemShut {NoStop}%
\bibitem [{\citenamefont {Carney}\ \emph {et~al.}(2019)\citenamefont {Carney},
  \citenamefont {Stamp},\ and\ \citenamefont {Taylor}}]{carney2018tabletop}%
  \BibitemOpen
  \bibfield  {author} {\bibinfo {author} {\bibfnamefont {D.}~\bibnamefont
  {Carney}}, \bibinfo {author} {\bibfnamefont {P.~C.~E.}\ \bibnamefont
  {Stamp}}, \ and\ \bibinfo {author} {\bibfnamefont {J.~M.}\ \bibnamefont
  {Taylor}},\ }\href {\doibase 10.1088/1361-6382/aaf9ca} {\bibfield  {journal}
  {\bibinfo  {journal} {Classical and Quantum Gravity}\ }\textbf {\bibinfo
  {volume} {36}},\ \bibinfo {pages} {034001} (\bibinfo {year}
  {2019})}\BibitemShut {NoStop}%
\bibitem [{\citenamefont {Anastopoulos}\ and\ \citenamefont
  {Hu}(2018)}]{anastopoulos2018comment}%
  \BibitemOpen
  \bibfield  {author} {\bibinfo {author} {\bibfnamefont {C.}~\bibnamefont
  {Anastopoulos}}\ and\ \bibinfo {author} {\bibfnamefont {B.-L.}\ \bibnamefont
  {Hu}},\ }\href@noop {} {\bibfield  {journal} {\bibinfo  {journal} {arXiv
  preprint arXiv:1804.11315}\ } (\bibinfo {year} {2018})}\BibitemShut {NoStop}%
\bibitem [{\citenamefont {Hall}\ and\ \citenamefont
  {Reginatto}(2018)}]{hall2018two}%
  \BibitemOpen
  \bibfield  {author} {\bibinfo {author} {\bibfnamefont {M.~J.}\ \bibnamefont
  {Hall}}\ and\ \bibinfo {author} {\bibfnamefont {M.}~\bibnamefont
  {Reginatto}},\ }\href@noop {} {\bibfield  {journal} {\bibinfo  {journal}
  {Journal of Physics A: Mathematical and Theoretical}\ }\textbf {\bibinfo
  {volume} {51}},\ \bibinfo {pages} {085303} (\bibinfo {year}
  {2018})}\BibitemShut {NoStop}%
\bibitem [{\citenamefont {Marletto}\ and\ \citenamefont
  {Vedral}(2018)}]{marletto2018can}%
  \BibitemOpen
  \bibfield  {author} {\bibinfo {author} {\bibfnamefont {C.}~\bibnamefont
  {Marletto}}\ and\ \bibinfo {author} {\bibfnamefont {V.}~\bibnamefont
  {Vedral}},\ }\href@noop {} {\bibfield  {journal} {\bibinfo  {journal}
  {Physical Review D}\ }\textbf {\bibinfo {volume} {98}},\ \bibinfo {pages}
  {046001} (\bibinfo {year} {2018})}\BibitemShut {NoStop}%
\bibitem [{\citenamefont {Christodoulou}\ and\ \citenamefont
  {Rovelli}(2018{\natexlab{a}})}]{superpositionofgeometries2018}%
  \BibitemOpen
  \bibfield  {author} {\bibinfo {author} {\bibfnamefont {M.}~\bibnamefont
  {Christodoulou}}\ and\ \bibinfo {author} {\bibfnamefont {C.}~\bibnamefont
  {Rovelli}},\ }\href@noop {} {\bibfield  {journal} {\bibinfo  {journal} {arXiv
  preprint arXiv:1808.05842}\ } (\bibinfo {year}
  {2018}{\natexlab{a}})}\BibitemShut {NoStop}%
\bibitem [{\citenamefont {Christodoulou}\ and\ \citenamefont
  {Rovelli}(2018{\natexlab{b}})}]{christodoulou2018possibility}%
  \BibitemOpen
  \bibfield  {author} {\bibinfo {author} {\bibfnamefont {M.}~\bibnamefont
  {Christodoulou}}\ and\ \bibinfo {author} {\bibfnamefont {C.}~\bibnamefont
  {Rovelli}},\ }\href@noop {} {\bibfield  {journal} {\bibinfo  {journal} {arXiv
  preprint arXiv:1812.01542}\ } (\bibinfo {year}
  {2018}{\natexlab{b}})}\BibitemShut {NoStop}%
\bibitem [{\citenamefont {Belenchia}\ \emph {et~al.}(2019)\citenamefont
  {Belenchia}, \citenamefont {Wald}, \citenamefont {Giacomini}, \citenamefont
  {Castro-Ruiz}, \citenamefont {Brukner},\ and\ \citenamefont
  {Aspelmeyer}}]{belenchia2019information}%
  \BibitemOpen
  \bibfield  {author} {\bibinfo {author} {\bibfnamefont {A.}~\bibnamefont
  {Belenchia}}, \bibinfo {author} {\bibfnamefont {R.~M.}\ \bibnamefont {Wald}},
  \bibinfo {author} {\bibfnamefont {F.}~\bibnamefont {Giacomini}}, \bibinfo
  {author} {\bibfnamefont {E.}~\bibnamefont {Castro-Ruiz}}, \bibinfo {author}
  {\bibfnamefont {{\v{C}}.}~\bibnamefont {Brukner}}, \ and\ \bibinfo {author}
  {\bibfnamefont {M.}~\bibnamefont {Aspelmeyer}},\ }\href@noop {} {\bibfield
  {journal} {\bibinfo  {journal} {arXiv preprint arXiv:1905.04496}\ } (\bibinfo
  {year} {2019})}\BibitemShut {NoStop}%
\bibitem [{\citenamefont {Giampaolo}\ and\ \citenamefont
  {Macr{\`\i}}(2018)}]{giampaolo2018entanglement}%
  \BibitemOpen
  \bibfield  {author} {\bibinfo {author} {\bibfnamefont {S.~M.}\ \bibnamefont
  {Giampaolo}}\ and\ \bibinfo {author} {\bibfnamefont {T.}~\bibnamefont
  {Macr{\`\i}}},\ }\href@noop {} {\bibfield  {journal} {\bibinfo  {journal}
  {arXiv preprint arXiv:1806.08383}\ } (\bibinfo {year} {2018})}\BibitemShut
  {NoStop}%
\bibitem [{\citenamefont {Carlesso}\ \emph {et~al.}(2019)\citenamefont
  {Carlesso}, \citenamefont {Bassi}, \citenamefont {Paternostro},\ and\
  \citenamefont {Ulbricht}}]{carlesso2019testing}%
  \BibitemOpen
  \bibfield  {author} {\bibinfo {author} {\bibfnamefont {M.}~\bibnamefont
  {Carlesso}}, \bibinfo {author} {\bibfnamefont {A.}~\bibnamefont {Bassi}},
  \bibinfo {author} {\bibfnamefont {M.}~\bibnamefont {Paternostro}}, \ and\
  \bibinfo {author} {\bibfnamefont {H.}~\bibnamefont {Ulbricht}},\ }\href@noop
  {} {\bibfield  {journal} {\bibinfo  {journal} {arXiv preprint
  arXiv:1906.04513}\ } (\bibinfo {year} {2019})}\BibitemShut {NoStop}%
\bibitem [{\citenamefont {Al~Balushi}\ \emph {et~al.}(2018)\citenamefont
  {Al~Balushi}, \citenamefont {Cong},\ and\ \citenamefont
  {Mann}}]{al2018optomechanical}%
  \BibitemOpen
  \bibfield  {author} {\bibinfo {author} {\bibfnamefont {A.}~\bibnamefont
  {Al~Balushi}}, \bibinfo {author} {\bibfnamefont {W.}~\bibnamefont {Cong}}, \
  and\ \bibinfo {author} {\bibfnamefont {R.~B.}\ \bibnamefont {Mann}},\
  }\href@noop {} {\bibfield  {journal} {\bibinfo  {journal} {Physical Review
  A}\ }\textbf {\bibinfo {volume} {98}},\ \bibinfo {pages} {043811} (\bibinfo
  {year} {2018})}\BibitemShut {NoStop}%
\bibitem [{\citenamefont {Page}\ and\ \citenamefont
  {Geilker}(1981)}]{Page:1981aj}%
  \BibitemOpen
  \bibfield  {author} {\bibinfo {author} {\bibfnamefont {D.~N.}\ \bibnamefont
  {Page}}\ and\ \bibinfo {author} {\bibfnamefont {C.}~\bibnamefont {Geilker}},\
  }\href {\doibase 10.1103/PhysRevLett.47.979} {\bibfield  {journal} {\bibinfo
  {journal} {Phys. Rev. Lett.}\ }\textbf {\bibinfo {volume} {47}},\ \bibinfo
  {pages} {979} (\bibinfo {year} {1981})}\BibitemShut {NoStop}%
\bibitem [{\citenamefont {Ashoorioon}\ \emph {et~al.}(2014)\citenamefont
  {Ashoorioon}, \citenamefont {Bhupal~Dev},\ and\ \citenamefont
  {Mazumdar}}]{Ashoorioon:2012kh}%
  \BibitemOpen
  \bibfield  {author} {\bibinfo {author} {\bibfnamefont {A.}~\bibnamefont
  {Ashoorioon}}, \bibinfo {author} {\bibfnamefont {P.}~\bibnamefont
  {Bhupal~Dev}}, \ and\ \bibinfo {author} {\bibfnamefont {A.}~\bibnamefont
  {Mazumdar}},\ }\href {\doibase 10.1142/S0217732314501636} {\bibfield
  {journal} {\bibinfo  {journal} {Mod. Phys. Lett. A}\ }\textbf {\bibinfo
  {volume} {29}},\ \bibinfo {pages} {1450163} (\bibinfo {year} {2014})},\
  \Eprint {http://arxiv.org/abs/1211.4678} {arXiv:1211.4678 [hep-th]}
  \BibitemShut {NoStop}%
\bibitem [{\citenamefont {Miao}\ \emph
  {et~al.}(2019{\natexlab{a}})\citenamefont {Miao}, \citenamefont {Martynov},\
  and\ \citenamefont {Yang}}]{Miao2019}%
  \BibitemOpen
  \bibfield  {author} {\bibinfo {author} {\bibfnamefont {H.}~\bibnamefont
  {Miao}}, \bibinfo {author} {\bibfnamefont {D.}~\bibnamefont {Martynov}}, \
  and\ \bibinfo {author} {\bibfnamefont {H.}~\bibnamefont {Yang}},\ }\href@noop
  {} {\bibfield  {journal} {\bibinfo  {journal} {arXiv preprint
  arXiv:1901.05827}\ } (\bibinfo {year} {2019}{\natexlab{a}})}\BibitemShut
  {NoStop}%
\bibitem [{\citenamefont {Bose}\ \emph {et~al.}(1997)\citenamefont {Bose},
  \citenamefont {Jacobs},\ and\ \citenamefont {Knight}}]{Bose1997}%
  \BibitemOpen
  \bibfield  {author} {\bibinfo {author} {\bibfnamefont {S.}~\bibnamefont
  {Bose}}, \bibinfo {author} {\bibfnamefont {K.}~\bibnamefont {Jacobs}}, \ and\
  \bibinfo {author} {\bibfnamefont {P.}~\bibnamefont {Knight}},\ }\href@noop {}
  {\bibfield  {journal} {\bibinfo  {journal} {Physical Review A}\ }\textbf
  {\bibinfo {volume} {56}},\ \bibinfo {pages} {4175} (\bibinfo {year}
  {1997})}\BibitemShut {NoStop}%
\bibitem [{\citenamefont {Bose}\ \emph {et~al.}(1999)\citenamefont {Bose},
  \citenamefont {Jacobs},\ and\ \citenamefont {Knight}}]{bose1999scheme}%
  \BibitemOpen
  \bibfield  {author} {\bibinfo {author} {\bibfnamefont {S.}~\bibnamefont
  {Bose}}, \bibinfo {author} {\bibfnamefont {K.}~\bibnamefont {Jacobs}}, \ and\
  \bibinfo {author} {\bibfnamefont {P.~L.}\ \bibnamefont {Knight}},\
  }\href@noop {} {\bibfield  {journal} {\bibinfo  {journal} {Physical Review
  A}\ }\textbf {\bibinfo {volume} {59}},\ \bibinfo {pages} {3204} (\bibinfo
  {year} {1999})}\BibitemShut {NoStop}%
\bibitem [{\citenamefont {Scala}\ \emph {et~al.}(2013)\citenamefont {Scala},
  \citenamefont {Kim}, \citenamefont {Morley}, \citenamefont {Barker},\ and\
  \citenamefont {Bose}}]{scala2013matter}%
  \BibitemOpen
  \bibfield  {author} {\bibinfo {author} {\bibfnamefont {M.}~\bibnamefont
  {Scala}}, \bibinfo {author} {\bibfnamefont {M.~S.}\ \bibnamefont {Kim}},
  \bibinfo {author} {\bibfnamefont {G.~W.}\ \bibnamefont {Morley}}, \bibinfo
  {author} {\bibfnamefont {P.~F.}\ \bibnamefont {Barker}}, \ and\ \bibinfo
  {author} {\bibfnamefont {S.}~\bibnamefont {Bose}},\ }\href@noop {} {\bibfield
   {journal} {\bibinfo  {journal} {Physical Review Letters}\ }\textbf {\bibinfo
  {volume} {111}},\ \bibinfo {pages} {180403} (\bibinfo {year}
  {2013})}\BibitemShut {NoStop}%
\bibitem [{\citenamefont {Yin}\ \emph {et~al.}(2013)\citenamefont {Yin},
  \citenamefont {Li}, \citenamefont {Zhang},\ and\ \citenamefont
  {Duan}}]{yin2013large}%
  \BibitemOpen
  \bibfield  {author} {\bibinfo {author} {\bibfnamefont {Z.~Q.}\ \bibnamefont
  {Yin}}, \bibinfo {author} {\bibfnamefont {T.}~\bibnamefont {Li}}, \bibinfo
  {author} {\bibfnamefont {X.}~\bibnamefont {Zhang}}, \ and\ \bibinfo {author}
  {\bibfnamefont {L.~M.}\ \bibnamefont {Duan}},\ }\href@noop {} {\bibfield
  {journal} {\bibinfo  {journal} {Physical Review A}\ }\textbf {\bibinfo
  {volume} {88}},\ \bibinfo {pages} {033614} (\bibinfo {year}
  {2013})}\BibitemShut {NoStop}%
\bibitem [{\citenamefont {Romero-Isart}(2011)}]{PhysRevA.84.052121}%
  \BibitemOpen
  \bibfield  {author} {\bibinfo {author} {\bibfnamefont {O.}~\bibnamefont
  {Romero-Isart}},\ }\href {\doibase 10.1103/PhysRevA.84.052121} {\bibfield
  {journal} {\bibinfo  {journal} {Phys. Rev. A}\ }\textbf {\bibinfo {volume}
  {84}},\ \bibinfo {pages} {052121} (\bibinfo {year} {2011})}\BibitemShut
  {NoStop}%
\bibitem [{\citenamefont {Bateman}\ \emph {et~al.}(2014)\citenamefont
  {Bateman}, \citenamefont {Nimmrichter}, \citenamefont {Hornberger},\ and\
  \citenamefont {Ulbricht}}]{bateman2014near}%
  \BibitemOpen
  \bibfield  {author} {\bibinfo {author} {\bibfnamefont {J.}~\bibnamefont
  {Bateman}}, \bibinfo {author} {\bibfnamefont {S.}~\bibnamefont
  {Nimmrichter}}, \bibinfo {author} {\bibfnamefont {K.}~\bibnamefont
  {Hornberger}}, \ and\ \bibinfo {author} {\bibfnamefont {H.}~\bibnamefont
  {Ulbricht}},\ }\href@noop {} {\bibfield  {journal} {\bibinfo  {journal}
  {Nature Communications}\ }\textbf {\bibinfo {volume} {5}},\ \bibinfo {pages}
  {4788} (\bibinfo {year} {2014})}\BibitemShut {NoStop}%
\bibitem [{\citenamefont {Marinkovi{\'c}}\ \emph {et~al.}(2018)\citenamefont
  {Marinkovi{\'c}}, \citenamefont {Wallucks}, \citenamefont {Riedinger},
  \citenamefont {Hong}, \citenamefont {Aspelmeyer},\ and\ \citenamefont
  {Gr{\"o}blacher}}]{marinkovic2018optomechanical}%
  \BibitemOpen
  \bibfield  {author} {\bibinfo {author} {\bibfnamefont {I.}~\bibnamefont
  {Marinkovi{\'c}}}, \bibinfo {author} {\bibfnamefont {A.}~\bibnamefont
  {Wallucks}}, \bibinfo {author} {\bibfnamefont {R.}~\bibnamefont {Riedinger}},
  \bibinfo {author} {\bibfnamefont {S.}~\bibnamefont {Hong}}, \bibinfo {author}
  {\bibfnamefont {M.}~\bibnamefont {Aspelmeyer}}, \ and\ \bibinfo {author}
  {\bibfnamefont {S.}~\bibnamefont {Gr{\"o}blacher}},\ }\href@noop {}
  {\bibfield  {journal} {\bibinfo  {journal} {Physical review letters}\
  }\textbf {\bibinfo {volume} {121}},\ \bibinfo {pages} {220404} (\bibinfo
  {year} {2018})}\BibitemShut {NoStop}%
\bibitem [{\citenamefont {Ahn}\ \emph {et~al.}(2018)\citenamefont {Ahn},
  \citenamefont {Xu}, \citenamefont {Bang}, \citenamefont {Deng}, \citenamefont
  {Hoang}, \citenamefont {Han}, \citenamefont {Ma},\ and\ \citenamefont
  {Li}}]{ahn2018optically}%
  \BibitemOpen
  \bibfield  {author} {\bibinfo {author} {\bibfnamefont {J.}~\bibnamefont
  {Ahn}}, \bibinfo {author} {\bibfnamefont {Z.}~\bibnamefont {Xu}}, \bibinfo
  {author} {\bibfnamefont {J.}~\bibnamefont {Bang}}, \bibinfo {author}
  {\bibfnamefont {Y.-H.}\ \bibnamefont {Deng}}, \bibinfo {author}
  {\bibfnamefont {T.~M.}\ \bibnamefont {Hoang}}, \bibinfo {author}
  {\bibfnamefont {Q.}~\bibnamefont {Han}}, \bibinfo {author} {\bibfnamefont
  {R.-M.}\ \bibnamefont {Ma}}, \ and\ \bibinfo {author} {\bibfnamefont
  {T.}~\bibnamefont {Li}},\ }\href@noop {} {\bibfield  {journal} {\bibinfo
  {journal} {Physical review letters}\ }\textbf {\bibinfo {volume} {121}},\
  \bibinfo {pages} {033603} (\bibinfo {year} {2018})}\BibitemShut {NoStop}%
\bibitem [{\citenamefont {Kaltenbaek}\ \emph {et~al.}(2016)\citenamefont
  {Kaltenbaek}, \citenamefont {Aspelmeyer}, \citenamefont {Barker},
  \citenamefont {Bassi}, \citenamefont {Bateman}, \citenamefont {Bongs},
  \citenamefont {Bose}, \citenamefont {Braxmaier}, \citenamefont {Brukner},
  \citenamefont {Christophe} \emph {et~al.}}]{kaltenbaek2016macroscopic}%
  \BibitemOpen
  \bibfield  {author} {\bibinfo {author} {\bibfnamefont {R.}~\bibnamefont
  {Kaltenbaek}}, \bibinfo {author} {\bibfnamefont {M.}~\bibnamefont
  {Aspelmeyer}}, \bibinfo {author} {\bibfnamefont {P.~F.}\ \bibnamefont
  {Barker}}, \bibinfo {author} {\bibfnamefont {A.}~\bibnamefont {Bassi}},
  \bibinfo {author} {\bibfnamefont {J.}~\bibnamefont {Bateman}}, \bibinfo
  {author} {\bibfnamefont {K.}~\bibnamefont {Bongs}}, \bibinfo {author}
  {\bibfnamefont {S.}~\bibnamefont {Bose}}, \bibinfo {author} {\bibfnamefont
  {C.}~\bibnamefont {Braxmaier}}, \bibinfo {author} {\bibfnamefont
  {{\v{C}}.}~\bibnamefont {Brukner}}, \bibinfo {author} {\bibfnamefont
  {B.}~\bibnamefont {Christophe}},  \emph {et~al.},\ }\href@noop {} {\bibfield
  {journal} {\bibinfo  {journal} {EPJ Quantum Technology}\ }\textbf {\bibinfo
  {volume} {3}},\ \bibinfo {pages} {5} (\bibinfo {year} {2016})}\BibitemShut
  {NoStop}%
\bibitem [{\citenamefont {Arndt}\ and\ \citenamefont
  {Hornberger}(2014)}]{arndt2014testing}%
  \BibitemOpen
  \bibfield  {author} {\bibinfo {author} {\bibfnamefont {M.}~\bibnamefont
  {Arndt}}\ and\ \bibinfo {author} {\bibfnamefont {K.}~\bibnamefont
  {Hornberger}},\ }\href@noop {} {\bibfield  {journal} {\bibinfo  {journal}
  {Nature Physics}\ }\textbf {\bibinfo {volume} {10}},\ \bibinfo {pages} {271}
  (\bibinfo {year} {2014})}\BibitemShut {NoStop}%
\bibitem [{\citenamefont {Miao}\ \emph
  {et~al.}(2019{\natexlab{b}})\citenamefont {Miao}, \citenamefont {Martynov},\
  and\ \citenamefont {Yang}}]{miao2019quantum}%
  \BibitemOpen
  \bibfield  {author} {\bibinfo {author} {\bibfnamefont {H.}~\bibnamefont
  {Miao}}, \bibinfo {author} {\bibfnamefont {D.}~\bibnamefont {Martynov}}, \
  and\ \bibinfo {author} {\bibfnamefont {H.}~\bibnamefont {Yang}},\ }\href@noop
  {} {\bibfield  {journal} {\bibinfo  {journal} {arXiv preprint
  arXiv:1901.05827}\ } (\bibinfo {year} {2019}{\natexlab{b}})}\BibitemShut
  {NoStop}%
\bibitem [{\citenamefont {Krisnanda}\ \emph {et~al.}(2019)\citenamefont
  {Krisnanda}, \citenamefont {Tham}, \citenamefont {Paternostro},\ and\
  \citenamefont {Paterek}}]{krisnanda2019observable}%
  \BibitemOpen
  \bibfield  {author} {\bibinfo {author} {\bibfnamefont {T.}~\bibnamefont
  {Krisnanda}}, \bibinfo {author} {\bibfnamefont {G.~Y.}\ \bibnamefont {Tham}},
  \bibinfo {author} {\bibfnamefont {M.}~\bibnamefont {Paternostro}}, \ and\
  \bibinfo {author} {\bibfnamefont {T.}~\bibnamefont {Paterek}},\ }\href@noop
  {} {\bibfield  {journal} {\bibinfo  {journal} {arXiv preprint
  arXiv:1906.08808}\ } (\bibinfo {year} {2019})}\BibitemShut {NoStop}%
\bibitem [{\citenamefont {Bykov}\ \emph {et~al.}(2019)\citenamefont {Bykov},
  \citenamefont {Mestres}, \citenamefont {Dania}, \citenamefont
  {Schm{\"o}ger},\ and\ \citenamefont {Northup}}]{bykov2019direct}%
  \BibitemOpen
  \bibfield  {author} {\bibinfo {author} {\bibfnamefont {D.~S.}\ \bibnamefont
  {Bykov}}, \bibinfo {author} {\bibfnamefont {P.}~\bibnamefont {Mestres}},
  \bibinfo {author} {\bibfnamefont {L.}~\bibnamefont {Dania}}, \bibinfo
  {author} {\bibfnamefont {L.}~\bibnamefont {Schm{\"o}ger}}, \ and\ \bibinfo
  {author} {\bibfnamefont {T.~E.}\ \bibnamefont {Northup}},\ }\href@noop {}
  {\bibfield  {journal} {\bibinfo  {journal} {arXiv preprint arXiv:1905.04204}\
  } (\bibinfo {year} {2019})}\BibitemShut {NoStop}%
\bibitem [{\citenamefont {Qvarfort}\ \emph {et~al.}(2018)\citenamefont
  {Qvarfort}, \citenamefont {Bose},\ and\ \citenamefont
  {Serafini}}]{qvarfort2018mesoscopic}%
  \BibitemOpen
  \bibfield  {author} {\bibinfo {author} {\bibfnamefont {S.}~\bibnamefont
  {Qvarfort}}, \bibinfo {author} {\bibfnamefont {S.}~\bibnamefont {Bose}}, \
  and\ \bibinfo {author} {\bibfnamefont {A.}~\bibnamefont {Serafini}},\
  }\href@noop {} {\bibfield  {journal} {\bibinfo  {journal} {arXiv preprint
  arXiv:1812.09776}\ } (\bibinfo {year} {2018})}\BibitemShut {NoStop}%
\bibitem [{\citenamefont {Wan}\ \emph {et~al.}(2016)\citenamefont {Wan},
  \citenamefont {Scala}, \citenamefont {Morley}, \citenamefont {Rahman},
  \citenamefont {Ulbricht}, \citenamefont {Bateman}, \citenamefont {Barker},
  \citenamefont {Bose},\ and\ \citenamefont {Kim}}]{PhysRevLett.117.143003}%
  \BibitemOpen
  \bibfield  {author} {\bibinfo {author} {\bibfnamefont {C.}~\bibnamefont
  {Wan}}, \bibinfo {author} {\bibfnamefont {M.}~\bibnamefont {Scala}}, \bibinfo
  {author} {\bibfnamefont {G.~W.}\ \bibnamefont {Morley}}, \bibinfo {author}
  {\bibfnamefont {A.~A.}\ \bibnamefont {Rahman}}, \bibinfo {author}
  {\bibfnamefont {H.}~\bibnamefont {Ulbricht}}, \bibinfo {author}
  {\bibfnamefont {J.}~\bibnamefont {Bateman}}, \bibinfo {author} {\bibfnamefont
  {P.~F.}\ \bibnamefont {Barker}}, \bibinfo {author} {\bibfnamefont
  {S.}~\bibnamefont {Bose}}, \ and\ \bibinfo {author} {\bibfnamefont {M.~S.}\
  \bibnamefont {Kim}},\ }\href {\doibase 10.1103/PhysRevLett.117.143003}
  {\bibfield  {journal} {\bibinfo  {journal} {Phys. Rev. Lett.}\ }\textbf
  {\bibinfo {volume} {117}},\ \bibinfo {pages} {143003} (\bibinfo {year}
  {2016})}\BibitemShut {NoStop}%
\bibitem [{\citenamefont {Machluf}\ \emph {et~al.}(2013)\citenamefont
  {Machluf}, \citenamefont {Japha},\ and\ \citenamefont {Folman}}]{Folman2013}%
  \BibitemOpen
  \bibfield  {author} {\bibinfo {author} {\bibfnamefont {S.}~\bibnamefont
  {Machluf}}, \bibinfo {author} {\bibfnamefont {Y.}~\bibnamefont {Japha}}, \
  and\ \bibinfo {author} {\bibfnamefont {R.}~\bibnamefont {Folman}},\ }\href
  {http://dx.doi.org/10.1038/ncomms3424} {\bibfield  {journal} {\bibinfo
  {journal} {Nature Communications}\ }\textbf {\bibinfo {volume} {4}},\
  \bibinfo {pages} {2424} (\bibinfo {year} {2013})}\BibitemShut {NoStop}%
\bibitem [{\citenamefont {Margalit}\ \emph {et~al.}(2018)\citenamefont
  {Margalit}, \citenamefont {Zhou}, \citenamefont {Dobkowski}, \citenamefont
  {Japha}, \citenamefont {Rohrlich}, \citenamefont {Moukouri},\ and\
  \citenamefont {Folman}}]{Folman2018}%
  \BibitemOpen
  \bibfield  {author} {\bibinfo {author} {\bibfnamefont {Y.}~\bibnamefont
  {Margalit}}, \bibinfo {author} {\bibfnamefont {Z.}~\bibnamefont {Zhou}},
  \bibinfo {author} {\bibfnamefont {O.}~\bibnamefont {Dobkowski}}, \bibinfo
  {author} {\bibfnamefont {Y.}~\bibnamefont {Japha}}, \bibinfo {author}
  {\bibfnamefont {D.}~\bibnamefont {Rohrlich}}, \bibinfo {author}
  {\bibfnamefont {S.}~\bibnamefont {Moukouri}}, \ and\ \bibinfo {author}
  {\bibfnamefont {R.}~\bibnamefont {Folman}},\ }\href@noop {} {\bibfield
  {journal} {\bibinfo  {journal} {arXiv preprint arXiv:1801.02708}\ } (\bibinfo
  {year} {2018})}\BibitemShut {NoStop}%
\bibitem [{\citenamefont {Amit}\ \emph {et~al.}(2019)\citenamefont {Amit},
  \citenamefont {Margalit}, \citenamefont {Dobkowski}, \citenamefont {Zhou},
  \citenamefont {Japha}, \citenamefont {Zimmermann}, \citenamefont {Efremov},
  \citenamefont {Narducci}, \citenamefont {Rasel}, \citenamefont {Schleich},\
  and\ \citenamefont {Folman}}]{folman2019}%
  \BibitemOpen
  \bibfield  {author} {\bibinfo {author} {\bibfnamefont {O.}~\bibnamefont
  {Amit}}, \bibinfo {author} {\bibfnamefont {Y.}~\bibnamefont {Margalit}},
  \bibinfo {author} {\bibfnamefont {O.}~\bibnamefont {Dobkowski}}, \bibinfo
  {author} {\bibfnamefont {Z.}~\bibnamefont {Zhou}}, \bibinfo {author}
  {\bibfnamefont {Y.}~\bibnamefont {Japha}}, \bibinfo {author} {\bibfnamefont
  {M.}~\bibnamefont {Zimmermann}}, \bibinfo {author} {\bibfnamefont {M.~A.}\
  \bibnamefont {Efremov}}, \bibinfo {author} {\bibfnamefont {F.~A.}\
  \bibnamefont {Narducci}}, \bibinfo {author} {\bibfnamefont {E.~M.}\
  \bibnamefont {Rasel}}, \bibinfo {author} {\bibfnamefont {W.~P.}\ \bibnamefont
  {Schleich}}, \ and\ \bibinfo {author} {\bibfnamefont {R.}~\bibnamefont
  {Folman}},\ }\href {\doibase 10.1103/PhysRevLett.123.083601} {\bibfield
  {journal} {\bibinfo  {journal} {Phys. Rev. Lett.}\ }\textbf {\bibinfo
  {volume} {123}},\ \bibinfo {pages} {083601} (\bibinfo {year}
  {2019})}\BibitemShut {NoStop}%
\bibitem [{\citenamefont {Bose}\ and\ \citenamefont
  {Morley}(2018)}]{Morley-Bose2018}%
  \BibitemOpen
  \bibfield  {author} {\bibinfo {author} {\bibfnamefont {S.}~\bibnamefont
  {Bose}}\ and\ \bibinfo {author} {\bibfnamefont {G.~W.}\ \bibnamefont
  {Morley}},\ }\href@noop {} {\bibfield  {journal} {\bibinfo  {journal} {arXiv
  preprint arXiv:1810.07045}\ } (\bibinfo {year} {2018})}\BibitemShut {NoStop}%
\bibitem [{\citenamefont {Pedernales}\ \emph {et~al.}(2019)\citenamefont
  {Pedernales}, \citenamefont {Morley},\ and\ \citenamefont
  {Plenio}}]{Pedernalles2019}%
  \BibitemOpen
  \bibfield  {author} {\bibinfo {author} {\bibfnamefont {J.~S.}\ \bibnamefont
  {Pedernales}}, \bibinfo {author} {\bibfnamefont {G.~W.}\ \bibnamefont
  {Morley}}, \ and\ \bibinfo {author} {\bibfnamefont {M.~B.}\ \bibnamefont
  {Plenio}},\ }\href@noop {} {\bibfield  {journal} {\bibinfo  {journal} {arXiv
  preprint arXiv:1906.00835}\ } (\bibinfo {year} {2019})}\BibitemShut {NoStop}%
\bibitem [{\citenamefont {Marshman}\ \emph {et~al.}(2018)\citenamefont
  {Marshman}, \citenamefont {Mazumdar}, \citenamefont {Morley}, \citenamefont
  {Barker}, \citenamefont {Hoekstra},\ and\ \citenamefont {Bose}}]{MIMAC2018}%
  \BibitemOpen
  \bibfield  {author} {\bibinfo {author} {\bibfnamefont {R.~J.}\ \bibnamefont
  {Marshman}}, \bibinfo {author} {\bibfnamefont {A.}~\bibnamefont {Mazumdar}},
  \bibinfo {author} {\bibfnamefont {G.~W.}\ \bibnamefont {Morley}}, \bibinfo
  {author} {\bibfnamefont {P.~F.}\ \bibnamefont {Barker}}, \bibinfo {author}
  {\bibfnamefont {S.}~\bibnamefont {Hoekstra}}, \ and\ \bibinfo {author}
  {\bibfnamefont {S.}~\bibnamefont {Bose}},\ }\href@noop {} {\bibfield
  {journal} {\bibinfo  {journal} {arXiv preprint arXiv:1807.10830}\ } (\bibinfo
  {year} {2018})}\BibitemShut {NoStop}%
\bibitem [{\citenamefont {Nguyen}\ and\ \citenamefont
  {Bernards}(2020)}]{nguyen2020entanglement}%
  \BibitemOpen
  \bibfield  {author} {\bibinfo {author} {\bibfnamefont {H.~C.}\ \bibnamefont
  {Nguyen}}\ and\ \bibinfo {author} {\bibfnamefont {F.}~\bibnamefont
  {Bernards}},\ }\href@noop {} {\bibfield  {journal} {\bibinfo  {journal} {The
  European Physical Journal D}\ }\textbf {\bibinfo {volume} {74}},\ \bibinfo
  {pages} {1} (\bibinfo {year} {2020})}\BibitemShut {NoStop}%
\bibitem [{\citenamefont {Geraci}\ \emph {et~al.}(2010)\citenamefont {Geraci},
  \citenamefont {Papp},\ and\ \citenamefont {Kitching}}]{Geraci2010}%
  \BibitemOpen
  \bibfield  {author} {\bibinfo {author} {\bibfnamefont {A.~A.}\ \bibnamefont
  {Geraci}}, \bibinfo {author} {\bibfnamefont {S.~B.}\ \bibnamefont {Papp}}, \
  and\ \bibinfo {author} {\bibfnamefont {J.}~\bibnamefont {Kitching}},\
  }\href@noop {} {\bibfield  {journal} {\bibinfo  {journal} {Physical review
  letters}\ }\textbf {\bibinfo {volume} {105}},\ \bibinfo {pages} {101101}
  (\bibinfo {year} {2010})}\BibitemShut {NoStop}%
\bibitem [{\citenamefont {{Pedernales}}\ \emph {et~al.}(2019)\citenamefont
  {{Pedernales}}, \citenamefont {{Morley}},\ and\ \citenamefont
  {{Plenio}}}]{2019arXiv190600835P}%
  \BibitemOpen
  \bibfield  {author} {\bibinfo {author} {\bibfnamefont {J.~S.}\ \bibnamefont
  {{Pedernales}}}, \bibinfo {author} {\bibfnamefont {G.~W.}\ \bibnamefont
  {{Morley}}}, \ and\ \bibinfo {author} {\bibfnamefont {M.~B.}\ \bibnamefont
  {{Plenio}}},\ }\href@noop {} {\bibfield  {journal} {\bibinfo  {journal}
  {arXiv e-prints}\ ,\ \bibinfo {eid} {arXiv:1906.00835}} (\bibinfo {year}
  {2019})},\ \Eprint {http://arxiv.org/abs/1906.00835} {arXiv:1906.00835
  [quant-ph]} \BibitemShut {NoStop}%
\bibitem [{\citenamefont {Edholm}\ \emph {et~al.}(2016)\citenamefont {Edholm},
  \citenamefont {Koshelev},\ and\ \citenamefont {Mazumdar}}]{Edholm:2016hbt}%
  \BibitemOpen
  \bibfield  {author} {\bibinfo {author} {\bibfnamefont {J.}~\bibnamefont
  {Edholm}}, \bibinfo {author} {\bibfnamefont {A.~S.}\ \bibnamefont
  {Koshelev}}, \ and\ \bibinfo {author} {\bibfnamefont {A.}~\bibnamefont
  {Mazumdar}},\ }\href {\doibase 10.1103/PhysRevD.94.104033} {\bibfield
  {journal} {\bibinfo  {journal} {Phys. Rev. D}\ }\textbf {\bibinfo {volume}
  {94}},\ \bibinfo {pages} {104033} (\bibinfo {year} {2016})},\ \Eprint
  {http://arxiv.org/abs/1604.01989} {arXiv:1604.01989 [gr-qc]} \BibitemShut
  {NoStop}%
\bibitem [{\citenamefont {Monteiro}\ \emph {et~al.}(2020)\citenamefont
  {Monteiro}, \citenamefont {Li}, \citenamefont {Afek}, \citenamefont {Li},
  \citenamefont {Mossman},\ and\ \citenamefont {Moore}}]{FMonteiro2020}%
  \BibitemOpen
  \bibfield  {author} {\bibinfo {author} {\bibfnamefont {F.}~\bibnamefont
  {Monteiro}}, \bibinfo {author} {\bibfnamefont {W.}~\bibnamefont {Li}},
  \bibinfo {author} {\bibfnamefont {G.}~\bibnamefont {Afek}}, \bibinfo {author}
  {\bibfnamefont {C.-l.}\ \bibnamefont {Li}}, \bibinfo {author} {\bibfnamefont
  {M.}~\bibnamefont {Mossman}}, \ and\ \bibinfo {author} {\bibfnamefont
  {D.~C.}\ \bibnamefont {Moore}},\ }\href@noop {} {\bibfield  {journal}
  {\bibinfo  {journal} {Physical Review A}\ }\textbf {\bibinfo {volume}
  {101}},\ \bibinfo {pages} {053835} (\bibinfo {year} {2020})}\BibitemShut
  {NoStop}%
\bibitem [{\citenamefont {Monteiro}\ \emph {et~al.}(2018)\citenamefont
  {Monteiro}, \citenamefont {Ghosh}, \citenamefont {van Assendelft},\ and\
  \citenamefont {Moore}}]{FMonteiro2018}%
  \BibitemOpen
  \bibfield  {author} {\bibinfo {author} {\bibfnamefont {F.}~\bibnamefont
  {Monteiro}}, \bibinfo {author} {\bibfnamefont {S.}~\bibnamefont {Ghosh}},
  \bibinfo {author} {\bibfnamefont {E.~C.}\ \bibnamefont {van Assendelft}}, \
  and\ \bibinfo {author} {\bibfnamefont {D.~C.}\ \bibnamefont {Moore}},\
  }\href@noop {} {\bibfield  {journal} {\bibinfo  {journal} {Physical Review
  A}\ }\textbf {\bibinfo {volume} {97}},\ \bibinfo {pages} {051802} (\bibinfo
  {year} {2018})}\BibitemShut {NoStop}%
\bibitem [{\citenamefont {Casimir}\ and\ \citenamefont
  {Polder}(1948)}]{Casmir:1947hx}%
  \BibitemOpen
  \bibfield  {author} {\bibinfo {author} {\bibfnamefont {H.}~\bibnamefont
  {Casimir}}\ and\ \bibinfo {author} {\bibfnamefont {D.}~\bibnamefont
  {Polder}},\ }\href {\doibase 10.1103/PhysRev.73.360} {\bibfield  {journal}
  {\bibinfo  {journal} {Phys. Rev.}\ }\textbf {\bibinfo {volume} {73}},\
  \bibinfo {pages} {360} (\bibinfo {year} {1948})}\BibitemShut {NoStop}%
\bibitem [{\citenamefont {Wan}(2017)}]{wan2017quantum}%
  \BibitemOpen
  \bibfield  {author} {\bibinfo {author} {\bibfnamefont {C.}~\bibnamefont
  {Wan}},\ }\href@noop {} {\emph {\bibinfo {title} {Quantum superposition on
  nano-mechanical oscillator}}}\ (\bibinfo  {publisher} {Imperial College
  London (PhD Thesis)},\ \bibinfo {year} {2017})\BibitemShut {NoStop}%
\bibitem [{\citenamefont {Kim}\ \emph {et~al.}(2005)\citenamefont {Kim},
  \citenamefont {Sofo}, \citenamefont {Velegol}, \citenamefont {Cole},
  \citenamefont {University},\ and\ \citenamefont
  {of~Technology-Bombay}}]{Kim2005StaticPO}%
  \BibitemOpen
  \bibfield  {author} {\bibinfo {author} {\bibfnamefont {H.-Y.}\ \bibnamefont
  {Kim}}, \bibinfo {author} {\bibfnamefont {J.~O.}\ \bibnamefont {Sofo}},
  \bibinfo {author} {\bibfnamefont {D.}~\bibnamefont {Velegol}}, \bibinfo
  {author} {\bibfnamefont {M.~W.}\ \bibnamefont {Cole}}, \bibinfo {author}
  {\bibfnamefont {G.~M. T. P.~S.}\ \bibnamefont {University}}, \ and\ \bibinfo
  {author} {\bibfnamefont {I.~I.}\ \bibnamefont {of~Technology-Bombay}},\
  }\href@noop {} {\bibfield  {journal} {\bibinfo  {journal} {Physical Review
  A}\ }\textbf {\bibinfo {volume} {72}},\ \bibinfo {pages} {053201} (\bibinfo
  {year} {2005})}\BibitemShut {NoStop}%
\bibitem [{\citenamefont {Ford}(1998)}]{Ford:1998ex}%
  \BibitemOpen
  \bibfield  {author} {\bibinfo {author} {\bibfnamefont {L.}~\bibnamefont
  {Ford}},\ }\href {\doibase 10.1103/PhysRevA.58.4279} {\bibfield  {journal}
  {\bibinfo  {journal} {Phys. Rev. A}\ }\textbf {\bibinfo {volume} {58}},\
  \bibinfo {pages} {4279} (\bibinfo {year} {1998})},\ \Eprint
  {http://arxiv.org/abs/quant-ph/9804055} {arXiv:quant-ph/9804055} \BibitemShut
  {NoStop}%
\bibitem [{\citenamefont {Ibarra}\ \emph {et~al.}(1997)\citenamefont {Ibarra},
  \citenamefont {Gonzalez}, \citenamefont {Vila},\ and\ \citenamefont
  {Molla}}]{article}%
  \BibitemOpen
  \bibfield  {author} {\bibinfo {author} {\bibfnamefont {A.}~\bibnamefont
  {Ibarra}}, \bibinfo {author} {\bibfnamefont {M.}~\bibnamefont {Gonzalez}},
  \bibinfo {author} {\bibfnamefont {R.}~\bibnamefont {Vila}}, \ and\ \bibinfo
  {author} {\bibfnamefont {J.}~\bibnamefont {Molla}},\ }\href {\doibase
  10.1016/S0925-9635(96)00724-8} {\bibfield  {journal} {\bibinfo  {journal}
  {Diamond and Related Materials}\ }\textbf {\bibinfo {volume} {6}},\ \bibinfo
  {pages} {856} (\bibinfo {year} {1997})}\BibitemShut {NoStop}%
\bibitem [{\citenamefont {Floch}\ \emph {et~al.}(2011)\citenamefont {Floch},
  \citenamefont {Bara}, \citenamefont {Hartnett}, \citenamefont {Tobar},
  \citenamefont {Mouneyrac}, \citenamefont {Passerieux}, \citenamefont {Cros},
  \citenamefont {Krupka}, \citenamefont {Goy},\ and\ \citenamefont
  {Caroopen}}]{Floch2011ElectromagneticPO}%
  \BibitemOpen
  \bibfield  {author} {\bibinfo {author} {\bibfnamefont {J.-M.~L.}\
  \bibnamefont {Floch}}, \bibinfo {author} {\bibfnamefont {R.}~\bibnamefont
  {Bara}}, \bibinfo {author} {\bibfnamefont {J.~G.}\ \bibnamefont {Hartnett}},
  \bibinfo {author} {\bibfnamefont {M.~E.}\ \bibnamefont {Tobar}}, \bibinfo
  {author} {\bibfnamefont {D.}~\bibnamefont {Mouneyrac}}, \bibinfo {author}
  {\bibfnamefont {D.}~\bibnamefont {Passerieux}}, \bibinfo {author}
  {\bibfnamefont {D.}~\bibnamefont {Cros}}, \bibinfo {author} {\bibfnamefont
  {J.}~\bibnamefont {Krupka}}, \bibinfo {author} {\bibfnamefont
  {P.}~\bibnamefont {Goy}}, \ and\ \bibinfo {author} {\bibfnamefont
  {S.}~\bibnamefont {Caroopen}}\ }(\bibinfo {year} {2011})\BibitemShut
  {NoStop}%
\bibitem [{\citenamefont {Ligatchev}(2008)}]{Ligatchev_2008}%
  \BibitemOpen
  \bibfield  {author} {\bibinfo {author} {\bibfnamefont {V.}~\bibnamefont
  {Ligatchev}},\ }in\ \href {\doibase 10.1149/1.2998530} {\emph {\bibinfo
  {booktitle} {{ECS} Transactions}}}\ (\bibinfo  {publisher} {{ECS}},\ \bibinfo
  {year} {2008})\BibitemShut {NoStop}%
\bibitem [{\citenamefont {Chevalier}\ \emph {et~al.}(2020)\citenamefont
  {Chevalier}, \citenamefont {Paige},\ and\ \citenamefont
  {Kim}}]{Chevalier:2020uvv}%
  \BibitemOpen
  \bibfield  {author} {\bibinfo {author} {\bibfnamefont {H.}~\bibnamefont
  {Chevalier}}, \bibinfo {author} {\bibfnamefont {A.}~\bibnamefont {Paige}}, \
  and\ \bibinfo {author} {\bibfnamefont {M.}~\bibnamefont {Kim}},\ }\href@noop
  {} {\  (\bibinfo {year} {2020})},\ \Eprint {http://arxiv.org/abs/2005.13922}
  {arXiv:2005.13922 [quant-ph]} \BibitemShut {NoStop}%
\bibitem [{\citenamefont {Schlosshauer}(2007)}]{decoherence}%
  \BibitemOpen
  \bibfield  {author} {\bibinfo {author} {\bibfnamefont {M.~A.}\ \bibnamefont
  {Schlosshauer}},\ }\href@noop {} {\emph {\bibinfo {title} {Decoherence and
  the Quantum-to Classical Transition}}}\ (\bibinfo  {publisher} {Springer},\
  \bibinfo {year} {2007})\BibitemShut {NoStop}%
\bibitem [{\citenamefont {Barry J.~Goodno}(2018)}]{book:2261446}%
  \BibitemOpen
  \bibfield  {author} {\bibinfo {author} {\bibfnamefont {J.~M.~G.}\
  \bibnamefont {Barry J.~Goodno}},\ }\href
  {http://gen.lib.rus.ec/book/index.php?md5=871c8a62c2d6039e58fff16898f16f05}
  {\emph {\bibinfo {title} {Mechanics of Materials}}},\ \bibinfo {edition}
  {9th}\ ed.,\ Mechanics of Materials (9e)\ (\bibinfo  {publisher} {Cengage
  Learning},\ \bibinfo {year} {2018})\BibitemShut {NoStop}%
\bibitem [{\citenamefont {Deli{\'c}}\ \emph {et~al.}(2020)\citenamefont
  {Deli{\'c}}, \citenamefont {Reisenbauer}, \citenamefont {Dare}, \citenamefont
  {Grass}, \citenamefont {Vuleti{\'c}}, \citenamefont {Kiesel},\ and\
  \citenamefont {Aspelmeyer}}]{Delic892}%
  \BibitemOpen
  \bibfield  {author} {\bibinfo {author} {\bibfnamefont {U.}~\bibnamefont
  {Deli{\'c}}}, \bibinfo {author} {\bibfnamefont {M.}~\bibnamefont
  {Reisenbauer}}, \bibinfo {author} {\bibfnamefont {K.}~\bibnamefont {Dare}},
  \bibinfo {author} {\bibfnamefont {D.}~\bibnamefont {Grass}}, \bibinfo
  {author} {\bibfnamefont {V.}~\bibnamefont {Vuleti{\'c}}}, \bibinfo {author}
  {\bibfnamefont {N.}~\bibnamefont {Kiesel}}, \ and\ \bibinfo {author}
  {\bibfnamefont {M.}~\bibnamefont {Aspelmeyer}},\ }\href {\doibase
  10.1126/science.aba3993} {\ \textbf {\bibinfo {volume} {367}},\ \bibinfo
  {pages} {892} (\bibinfo {year} {2020})}\BibitemShut {NoStop}%
\bibitem [{\citenamefont {Meyer}\ \emph {et~al.}(2019)\citenamefont {Meyer},
  \citenamefont {Sommer}, \citenamefont {Mestres}, \citenamefont {Gieseler},
  \citenamefont {Jain}, \citenamefont {Novotny},\ and\ \citenamefont
  {Quidant}}]{PhysRevLett.123.153601}%
  \BibitemOpen
  \bibfield  {author} {\bibinfo {author} {\bibfnamefont {N.}~\bibnamefont
  {Meyer}}, \bibinfo {author} {\bibfnamefont {A.~d. l.~R.}\ \bibnamefont
  {Sommer}}, \bibinfo {author} {\bibfnamefont {P.}~\bibnamefont {Mestres}},
  \bibinfo {author} {\bibfnamefont {J.}~\bibnamefont {Gieseler}}, \bibinfo
  {author} {\bibfnamefont {V.}~\bibnamefont {Jain}}, \bibinfo {author}
  {\bibfnamefont {L.}~\bibnamefont {Novotny}}, \ and\ \bibinfo {author}
  {\bibfnamefont {R.}~\bibnamefont {Quidant}},\ }\href {\doibase
  10.1103/PhysRevLett.123.153601} {\bibfield  {journal} {\bibinfo  {journal}
  {Phys. Rev. Lett.}\ }\textbf {\bibinfo {volume} {123}},\ \bibinfo {pages}
  {153601} (\bibinfo {year} {2019})}\BibitemShut {NoStop}%
\bibitem [{\citenamefont {Dwyer}\ \emph {et~al.}(2014)\citenamefont {Dwyer},
  \citenamefont {Pomeroy}, \citenamefont {Simons}, \citenamefont {Steffens},\
  and\ \citenamefont {Lau}}]{dwyer2014enriching}%
  \BibitemOpen
  \bibfield  {author} {\bibinfo {author} {\bibfnamefont {K.~J.}\ \bibnamefont
  {Dwyer}}, \bibinfo {author} {\bibfnamefont {J.~M.}\ \bibnamefont {Pomeroy}},
  \bibinfo {author} {\bibfnamefont {D.~S.}\ \bibnamefont {Simons}}, \bibinfo
  {author} {\bibfnamefont {K.}~\bibnamefont {Steffens}}, \ and\ \bibinfo
  {author} {\bibfnamefont {J.}~\bibnamefont {Lau}},\ }\href@noop {} {\bibfield
  {journal} {\bibinfo  {journal} {Journal of Physics D: Applied Physics}\
  }\textbf {\bibinfo {volume} {47}},\ \bibinfo {pages} {345105} (\bibinfo
  {year} {2014})}\BibitemShut {NoStop}%
\bibitem [{\citenamefont {Maldacena}(1999)}]{Maldacena:1997re}%
  \BibitemOpen
  \bibfield  {author} {\bibinfo {author} {\bibfnamefont {J.~M.}\ \bibnamefont
  {Maldacena}},\ }\href {\doibase 10.1023/A:1026654312961} {\bibfield
  {journal} {\bibinfo  {journal} {Int. J. Theor. Phys.}\ }\textbf {\bibinfo
  {volume} {38}},\ \bibinfo {pages} {1113} (\bibinfo {year} {1999})},\ \Eprint
  {http://arxiv.org/abs/hep-th/9711200} {arXiv:hep-th/9711200} \BibitemShut
  {NoStop}%
\end{thebibliography}%

\appendix

\section{Phase evolutions step 1 to step 3}\label{sec:extra phase}

The initial separation of the test-mass is  $d$. During the acceleration period of step-1 the superposition is created due to the acceleration created by the magnetic field gradient. The phase evolution of  Eq.(\ref{eq:entanglement_phase}) during this acceleration period of $\tau/2$ will be given by:
\begin{widetext}
	\begin{align}
	\Phi_{\text{eff}}=  \displaystyle \frac{Gm^2}{\hbar}\int_0^{\frac{\tau}{2}}\left(\frac{1}{d-a_{mag}t^2}+\frac{1}{d+a_{mag}t^2}-\frac{2}{d}\right)\text{d}t 
	= \displaystyle \frac{Gm^2}{\hbar}\left( -\frac{\ln\left(\frac{\left|\frac{a_{mag}\tau}{2}-\sqrt{a_{mag}d}\right|}{\frac{a_{mag}\tau}{2}+\sqrt{a_{mag}d}}\right)}{2\sqrt{a_{mag}d}}+\frac{\arctan\left(\frac{a_{mag}\tau}{2\sqrt{a_{mag}d}}\right)}{\sqrt{a_{mag}d}}-\frac{\tau}{d}\right)
	\label{eq: phase step 1 1}
	\end{align}
\end{widetext}
During the deceleration period, the phase evolution is given by:
\begin{widetext}
	\begin{align}
	\Phi_{\text{eff}}= \frac{Gm^2}{\hbar}\int_0^{\frac{\tau}{2}}\bigg(\frac{1}{d+a_{mag}t^2-a_{mag}\tau t -a_{mag}\left(\frac{\tau}{2}\right)^2} 
	+\frac{1}{d-a_{mag}t^2+a_{mag}\tau t  +a_{mag}\left(\frac{\tau}{2}\right)^2}-\frac{2}{d}\bigg)\textrm{d}t 
	\end{align}
\end{widetext}
Now by defining $d_1=d-a_{mag}\left(\frac{\tau}{2}\right)^2$ and $d_2=d+a_{mag} \left(\frac{\tau}{2}\right)^2$, we arrive at the following solution:
\begin{widetext}
	\begin{align}
	\Phi_{\text{eff}}=&\frac{Gm^2}{\hbar}\bigg( \frac{2}{\sqrt{4d_1a_{mag}-\left(a\tau\right)^2}}\textrm{arctan}\left(\frac{a_{mag}\tau}{\sqrt{4d_1a_{mag}-\left(a_{mag}\tau\right)^2}}\right) \nonumber\\
	&-\frac{1}{\sqrt{\left(a_{mag}\tau\right)^2+4a_{mag}d_2}}\text{ln}\left(\frac{-a_{mag}\tau+\sqrt{\left(a_{mag}\tau\right)^2+4a_{mag}d_2}}{a_{mag}\tau+\sqrt{\left(a_{mag}\tau\right)^2+4a_{mag}d_2}}\right) 
	-\frac{\tau}{d}\bigg)\label{eq: phase step 1 2}
	\end{align}
\end{widetext}
Where the assumptions are made that $4d_1a_{mag}>\left(a_{mag}\tau\right)^2$ and $\left(a_{mag}\tau\right)^2<4a_{mag}d_2$, which hold for the parameters of $m=10^{-15}$ kg discussed in this paper. Therefore  Eqs.\ref{eq: phase step 1 1} and \ref{eq: phase step 1 2} describe the total phase evolution during step-1 of the experiment.

In step-3 the superposition size is slightly enhanced due to Casimir force which tends to attract the nearby states to the plate, $\Delta x + s_{\text{max}}$, with $s_{\text{max}}$ the value of s at the end of the flight time, but as $\Delta x$ is an order of magnitude larger it is fairly reasonable to make the assumption that the phase evolution of step-3 is the same as that of step-1 with $\tau \rightarrow \tau_1$ and $d \rightarrow d-s_{\text{max}}$. With $\tau_1$ and $d'$ the assumptions $4d_1a>\left(a_{mag}\tau\right)^2$ and $\left(a\tau\right)^2<4a_{mag}d_2$ still hold. Therefore the total phase evolution during step 1 and step 3 of the experiment are given by Eqs.\ref{eq: phase step 1 1} and \ref{eq: phase step 1 2} for step-1 and the same equations with $\tau \rightarrow \tau_1$ and $d \rightarrow d-s_{\text{max}}$ for step-3.


\section{Scattering constants of decoherence rates \label{sec: scattering constants}}

The Ref.\cite{decoherence} has a detailed analysis on the decoherence of a spatially separated superposition of two macroscopic states like in in our case. The off-diagonal elements of the reduced density matrix $\rho$ of  a single superposition evolves due to loss of coherence due to interaction with environmental particles. The master equation is given by: 
\begin{equation}
\frac{\partial \rho(\textbf{x},\textbf{x'},t)}{\partial t}=-F(\textbf{x}-\textbf{x'})\rho(\textbf{x},\textbf{x'},t)\label{eq:reduced_evolution}
\end{equation}
Where $(\textbf{x}-\textbf{x'})$ is the superposition size.
If the wavelength of the environmental particles/photons are much shorter than the superposition size we have the so called short wavelength limit in which it can be shown \cite{decoherence} that for an isotropic medium $F(\textbf{x}-\textbf{x'})=\Gamma$, with $\Gamma$ the scattering rate of the environmental particles/photons independent of time, in an adiabatic limit.

If the wavelength of the environmental particles/photons is much larger than the superposition size, we are in the long wavelength limit, in which case $F(\textbf{x}-\textbf{x'})=\Lambda(\textbf{x}-\textbf{x'})^2$ \cite{decoherence}. 
However note that in our case, $(\textbf{x}-\textbf{x'})$, is time-dependent during the whole experiment. As mentioned in section \ref{sec:col decoherence} we have scattering of air molecules valid for a short wavelength limit and photons from absorption and scattering from the environment and emission from the test-mass for the long wavelength limit. This will yield that  Eq.(\ref{eq:reduced_evolution}) reduces to:
\begin{equation}
\frac{\partial \rho(\textbf{x},\textbf{x'},t)}{\partial t}=(\Gamma_{air}+\sum_i^3\Lambda_i(\textbf{x}(t)-\textbf{x'}(t))^2\rho(\textbf{x},\textbf{x'},t)\,,
\end{equation}
And
\begin{equation}
\rho(\textbf{x},\textbf{x'},t)=\text{exp}[-(\Gamma_{air}t+\sum_i^3\int( \textbf{x}(t)-\textbf{x'}(t))^2\text{d}t) ]\rho(\textbf{x},\textbf{x'},0)\,.
\end{equation}
In the small time interval limit, this integration can be expressed by a summation with $\Delta t$ taken sufficiently small, and therefore we can obtain the total decoherence factor $e^{-\gamma t}$:
\begin{equation}
\text{exp}[-\gamma t]=\text{exp}[-(\Gamma_{air}t+\sum_i^3\Lambda_i\sum_k( \textbf{x}_k-\textbf{x'}_k)^2\Delta t ]\,,
\end{equation}
here t denotes the final time of evolution of the summation given by $k_{max}\Delta t$. In our case it is the total time of steps-1,2 and 3 of the experiment.
In the long-wavelength limit the scattering constant for air molecules can be expressed as~\cite{decoherence}:
\begin{equation}
\Lambda_{\text{air}}=\frac{4R^2}{3\hbar^2}\frac{N}{V}\sqrt{\pi m_{air}}\left(2k_bT\right)^{3/2}\,,
\label{eq: scatter constant air}
\end{equation}
Where $k_b$ is the Boltzmann constant.
Note that $\Gamma_{air}=\lambda_{air}^2\Lambda_{air}$, see \cite{PhysRevA.84.052121}, and denoting $\frac{N}{V}=n_V$ we will obtain:
\begin{equation}
\Gamma_{\text{air}}=\frac{16\pi n_VR^2}{3}\sqrt{\frac{2\pi k_bT_{ex}}{{m_{\text{air}}}}} \label{eq: scattering rate air}
\end{equation}
Which is consistent with Ref.\cite{PhysRevA.84.052121} if one applies the ideal gas law, and $T_{ex}$ is the external temperature of the ambience.

The relevant formulas for the scattering constants for the photon scattering, absorption and emission can be found in \cite{decoherence, PhysRevA.84.052121}, and are given by:
\begin{align}
\Lambda_{sc}=&8!\zeta(9)\frac{8cR^6}{9\pi}\left( \frac{k_b T_{ex}}{\hbar c}\right)^9\text{Re}\left(\frac{\epsilon-1}{\epsilon+2}\right)^2 \nonumber \\
\Lambda_{(e)a}=&\frac{16\pi^5cR^3}{189}\left(\frac{k_bT_{(i)ex}}{\hbar c}\right)^6\text{Im}\left(\frac{\epsilon-1}{\epsilon+2}\right) \label{eq:Lambda_s_e_a}
\end{align}
Here $T_i$ is the internal temperature of the test-mass.


\section{Computing the decoherence rate \label{sec:decoherence rate}}

Note that from  Eq.(\ref{eq:decoherence factor})
$x_{\ket{\uparrow}_k}-x_{\ket{\downarrow}_k}$ is time-dependent. Step-1 of Fig.1 can be split up into an acceleration and deceleration period of the superposition size \cite{Bose:2017nin}. For the acceleration part, we get with the constant acceleration of $a_{mag}=\frac{g\mu_B\partial_xB}{m}$ during a time of $\frac{\tau}{2}$ that the summation of Eq.(\ref{eq:decoherence factor}) becomes:
\begin{equation}
\sum_k(x_{\ket{\uparrow}_k}-x_{\ket{\downarrow}_k})^2\Delta t=a_{mag}^2\int_0^{\tau/2}t^4\text{d}t=\frac{a_{mag}^2}{5}\left(\frac{\tau}{2}\right)^5  \label{eq:sum_k 1}
\end{equation}
We can perform the same analysis for the deceleration period, 
\begin{widetext}
	\begin{align}
	\sum_k(x_{\ket{\uparrow}_k}-x_{\ket{\downarrow}_k})^2\Delta t= \int_0^{\tau/2}\bigg(-a_{mag}t^2+2a_{mag}\frac{\tau}{2}t 
	+ a_{mag}\left(\frac{\tau}{2}\right)^2\bigg)^2\text{d}t
	= -\frac{a_{mag}^2}{5}\left(\frac{\tau}{2}\right)^5+\frac{46}{15}a_{mag}^2\left(\frac{\tau}{2}\right)^5 \label{eq:sum_k 2}
	\end{align}
\end{widetext}
Combining both the expressions will give us that during step-1 of the experiment:
\begin{equation}
\sum_k(x_{\ket{\uparrow}_k}-x_{\ket{\downarrow}_k})^2\Delta t= \frac{46}{15}a_{mag}^2\left(\frac{\tau}{2}\right)^5
\label{eq:sum_k 3}
\end{equation}
During the free-fall period of the experiment, see  Fig.\ref{fig:set-up2}, $x_{\ket{\uparrow}_k}-x_{\ket{\downarrow}_k}=\Delta x+s_k$ and by using that $\Delta x=2a_{mag}\left(\frac{\tau}{2}\right)^2$,  we obtain:
\begin{widetext}
	\begin{align}
	\sum_{i=1}^3\Lambda_i\sum_k(x_{\ket{\uparrow}_k}-x_{\ket{\downarrow}_k})^2\Delta t=& \sum_{i=1}^3\Lambda_i\sum_k(\Delta x +s_k)^2\Delta t \nonumber\\
	=&  \sum_{i=1}^3\Lambda_i[4a_{mag}^2\left(\frac{\tau}{2}\right)^4t_{int}+\sum_k(4a_{mag}\left(\frac{\tau}{2}\right)^2 s_k+s_k^2)\Delta t] \label{eq: 4}
	\end{align}
\end{widetext}
Finally, at the end of the free-fall we we will have the superposition size $\Delta$x+$s_{\text{max}}$. Then we  make  the assumption that the inner states in Fig.2 instantaneously turn around when the magnetic field is switched back on. However in reality this will take longer due to residual velocity of the state towards the plate, but this will for the mass of $10^{-15}$kg with the parameters of the interaction time of 1s happen after an aditional time of 10 ms and can even be reduced if one plays around with slightly higher magnetic field gradients during step-3 of the experiment, or taking slightly higher values of N. In short, there are a lot of possibilities. With this assumption, we get that the time required to bring the superposition back becomes
\begin{equation}
\tau_1=2\sqrt{\left(\frac{\tau}{2}\right)^2+\frac{s_{max}}{a_{mag}}}
\end{equation}
As step-3 of the experiment is the reverse action of step-1, but with now a time of $\tau_1$ instead of $\tau$, we obtain the summation, equivalent to Eq.(\ref{eq:sum_k 3}), but with $\tau \rightarrow \tau_1$:
\begin{equation}
\sum_k(x_{\ket{\uparrow}_k}-x_{\ket{\downarrow}_k})^2\Delta t= \frac{46}{15}a_{mag}^2\left(\frac{\tau_1}{2}\right)^5
\label{eq:sum_k 4}
\end{equation}
Then by combining expressions \ref{eq:sum_k 3},\ref{eq: 4} and \ref{eq:sum_k 4}, and denoting that the total runtime of the superposition is $t_{int}+\tau +\tau_1$, we will get that the total decoherence factor becomes:
\begin{widetext}
	\begin{align}
	\text{exp}[-\sum_k\gamma_k\Delta t]=&  \text{exp}[-[\Gamma_{\text{air}}(t_{\text{int}}+\tau+\tau_1 ) 
	+  \sum_{i=1}^3\Lambda_i\bigg(\frac{46}{15}a_{mag}^2\{\left(\frac{\tau}{2}\right)^5 +\left(\frac{\tau_1}{2}\right)^5\}+4a_{mag}^2\left(\frac{\tau}{2}\right)^4t_{int} 
	+  \sum_k(4a_{mag}\left(\frac{\tau}{2}\right)^2 s_k+s_k^2)\Delta t\bigg)]] \label{eq:full decohere analytic2 appendix}
	\end{align}
\end{widetext}
Where we have switched form the notation $\gamma t$ to $\sum_k\gamma_k\Delta t$ as the decoherence rate is not constant during the altered \text{QGEM} protocol.

\end{document}